%% file: lfi_instrument_description.tex
\documentclass[longauth,traditabstract]{aa}

\usepackage{graphicx}
\usepackage[dvips]{color} 
\usepackage{epsfig}
\usepackage{amssymb,amsmath}
\usepackage{longtable}
\usepackage{natbib}
\bibpunct{(}{)}{;}{a}{}{,} 

\def\sup#1{$^{\rm #1}$}
\def\lsim{\mathrel{\raise .4ex\hbox{\rlap{$<$}\lower 1.2ex\hbox{$\sim$}}}}
\def\gsim{\mathrel{\raise .4ex\hbox{\rlap{$>$}\lower 1.2ex\hbox{$\sim$}}}}
\def\deg{\ifmmode^\circ\else$^\circ$\fi}
\def\pdeg{\ifmmode $\setbox0=\hbox{$^{\circ}$}\rlap{\hskip.11\wd0 .}$^{\circ}
          \else \setbox0=\hbox{$^{\circ}$}\rlap{\hskip.11\wd0 .}$^{\circ}$\fi}
\def\arcs{\ifmmode {^{\scriptscriptstyle\prime\prime}}
          \else $^{\scriptscriptstyle\prime\prime}$\fi}
\def\arcm{\ifmmode {^{\scriptscriptstyle\prime}}
          \else $^{\scriptscriptstyle\prime}$\fi}
\newdimen\sa  \newdimen\sb
\def\parcs{\sa=.07em \sb=.03em
     \ifmmode \hbox{\rlap{.}}^{\scriptscriptstyle\prime\kern -\sb\prime}\hbox{\kern -\sa}
     \else \rlap{.}$^{\scriptscriptstyle\prime\kern -\sb\prime}$\kern -\sa\fi}
\def\parcm{\sa=.08em \sb=.03em
     \ifmmode \hbox{\rlap{.}\kern\sa}^{\scriptscriptstyle\prime}\hbox{\kern-\sb}
     \else \rlap{.}\kern\sa$^{\scriptscriptstyle\prime}$\kern-\sb\fi}
\def\,{\thinspace}
\newbox\tablebox    \newdimen\tablewidth
\def\leaderfil{\leaders\hbox to 5pt{\hss.\hss}\hfil}

\def\tabl #1\par{\centerline{TABLE \number\tableno}
                 \vskip 7pt
                 \centerline{#1}\nointerlineskip\noindent}
\def\enndtable{\tablewidth=\wd\tablebox 
    $$\hss\copy\tablebox\hss$$
    \vskip-\lastskip\vskip -2pt}
\def\tablenote#1 #2\par{\begingroup \parindent=0.8em
    \abovedisplayshortskip=0pt\belowdisplayshortskip=0pt
    \noindent
    $$\hss\vbox{\hsize\tablewidth \hangindent=\parindent \hangafter=1 \noindent
    \hbox to \parindent{\sup{\rm #1}\hss}\strut#2\strut\par}\hss$$
    \endgroup}
\def\doubleline{\vskip 3pt\hrule \vskip 1.5pt \hrule \vskip 5pt}
\def\muKs{\ifmmode \,\mu$K\,s$^{1/2}
          \else \,$\mu$K\,s$^{1/2}$\fi}
\def\lea{\mathrel{\raise .4ex\hbox{\rlap{$<$}\lower 1.2ex\hbox{$\sim$}}}}
\def\gea{\mathrel{\raise .4ex\hbox{\rlap{$>$}\lower 1.2ex\hbox{$\sim$}}}}
\let\lsim=\lea
\let\gsim=\gea
\def\mo{\ifmmode ^{-1}\else $^{-1}$\fi}
\def\muKs{\ifmmode \,\mu$K\,s$^{1/2}
           \else \,$\mu$K\,s$^{1/2}$\fi}
\def\microns{\ifmmode \,\mu$m$\else \,$\mu$m\fi}

\let\tfs=\scriptsize

\begin{document}

\input{00_title}

\maketitle

%

\def\aap{A\&A}%
\def\aapr{A\&A~Rev.}%
\def\aaps{A\&AS}%

\section{Introduction}
\label{sec:introduction}

  \input{100_introduction}


\section{Scientific requirements}
\label{sec:scientific_requirements}

 \input{200_scientific_requirements}

    \subsection{Frequency range}
    \label{sec:frequency_range}
    
        \input{210_frequency_range}

    \subsection{Angular resolution and sensitivity}
    \label{sec:sensitivity_angular_resolution}
    
        \input{220_angular_resolution_sensitivity}

        \subsubsection{Angular resolution}
        \label{sec:angular_resolution}

            \input{221_angular_resolution}
        \subsubsection{Sensitivity}
        \label{sec:sensitivity}

            \input{222_sensitivity}

    \subsection{Sensitivity budget}
    \label{sec:sensitivity_budget}
    
        \input{230_sensitivity_budget}

        \subsubsection{Bandwidth and system noise}
        \label{sec:bw_tnoise}

            \input{231_bandwidth_system_noise}

        \subsubsection{Active cooling}
        \label{sec:active_cooling}
        
            \input{232_active_cooling}

        \subsubsection{Breakdown allocations}
        \label{sec:breakdown_allocations}

            \input{233_breakdown_allocations}

    \subsection{Stability}
    \label{sec:stability}

        \input{240_stability}

    \subsection{Systematic effects}
    \label{sec:systematics}

        \input{250_systematics_requirements}


\section{Instrument concept}
\label{sec:instrument_concept}

    \input{300_instrument_concept}

    \subsection{Radiometer Chain Assemblies}
    \label{sec:rcas}

        \input{310_rcas}

    \subsection{Radiometer Array Assembly}
    \label{sec:raa}

        \input{320_raa}

    \subsection{Receiver design}
    \label{sec:receiver_design}

        \input{330_receiver_design}

        \subsubsection{Signal model}
        \label{sec:signal_model}

            \input{331_signal_model}

        \subsubsection{Knee frequency and gain modulation factor}
        \label{sec:fk_and_gmf}

            \input{332_knee_frequency_and_gmf}
        \subsubsection{Noise temperature}
        \label{sec:tnoise}

            \input{333_noise_temperature}

%

\section{LFI configuration and subsystems}
\label{sec:lfi_configuration_subsystems}

    \input{400_configuration_subsystems}
    \subsection{The Front-End Unit}
    \label{sec:feu}
    
        \subsubsection{Focal plane design}
        \label{sec:focal_plane_design}
            \input{411_focal_plane_design}

        \subsubsection{LFI main frame}
        \label{sec:main_frame}

            \input{412_lfi_main_frame}

        \subsubsection{Feed horns}
        \label{sec:feeds}
        
            \input{413_feed_horns}

        \subsubsection{Orthomode transducers}
        \label{sec:omts}

            \input{414_omts}

        \subsubsection{Front-end modules}
        \label{sec:fems}

            \input{415_fems}

    \subsection{The 4~K reference load system}
    \label{sec:4krl}

\input{420_the_rl_system}

    \subsection{Waveguides}
    \label{sec:wgs}

        \input{430_wgs}

    \subsection{Back-end unit}
    \label{sec:beu}
        
        \subsubsection{Back-end modules}
        \label{sec:bems}

            \input{441_bems}

        \subsubsection{Data acquisition electronics}
        \label{sec:dae}

            \input{442_dae}

        \subsection{Radiometer electronics box (REBA)}
        \label{sec:reba}

            \input{450_reba}


\section{Thermal interfaces}
\label{sec:thermal_interfaces}

    \subsection{LFI 20~K stage}
    \label{sec:20k_stage}
    
        \input{510_20k_stage}

    \subsection{Thermal loads}
    \label{sec:thermal_loads}

        \input{520_thermal_loads}

    \subsection{Temperature sensors}
    \label{sec:temperature_sensors}

        \input{530_temperature_sensors}


\section{Electrical and communication interfaces}
\label{sec:electrical_interfaces}

    \subsection{Cryo-harness}
    \label{sec:cryoharness}

\input{610_cryo_harness}

    \subsection{Electromagnetic compatibility (EMC)}
    \label{sec:emc}

        \input{620_emc}

    \subsection{Data rate}
    \label{sec:data_rate}

        \input{630_data_rate}


\section{Optical interfaces}
\label{sec:optical_interfaces}

    \input{700_optical_interfaces}

%
%
%

\section{Conclusions}
\label{sec:conclusions}

  \input{900_conclusions}

\begin{acknowledgements}

The Planck-LFI project is developed by an International Consortium lead by Italy and involving Canada, Finland, Germany, Norway, Spain, Switzerland, UK, USA. The Italian contribution to Planck is supported by the Italian Space Agency (ASI). TP's work was supported in part by the Academy of Finland grants 205800, 214598, 121703, and 121962. TP thanks Waldemar von Frenckells stiftelse, Magnus Ehrnrooth Foundation, and V\"ais\"al\"a Foundation for financial support. We acknowledge partial support of the NASA LTSA Grant NNG04CG90G.

\end{acknowledgements}

%



%
\bibliographystyle{aa}
\bibliography{./referencesmio}
\end{document}

%% file: 00_title.tex
\title{Planck pre-launch status: Design and description of the Low Frequency Instrument}

   \author{
 M.~Bersanelli\inst{1}$^,$\inst{2}\and
 N.~Mandolesi\inst{3}\and
 R.~C.~Butler\inst{3}\and
 A.~Mennella\inst{1}$^,$\inst{2}\and
 F.~Villa\inst{3}\and
 B.~Aja\inst{4}\and
 E.~Artal\inst{4}\and
 E.~Artina\inst{5}\and
 C.~Baccigalupi\inst{6}$^,$\inst{15}\and
 M.~Balasini\inst{5}
% ...
}

\offprints{M. Bersanelli}

\institute{
    Universit\`a degli Studi di Milano, Dipartimento di Fisica, via Celoria 16, 20133 Milano, Italy (e-mail: marco.bersanelli@unimi.it) \and
    INAF -- Istituto di Astrofisica Spaziale e Fisica Cosmica, via  Bassini 15, 20133 Milano, Italy\and
    INAF -- Istituto di Astrofisica Spaziale e Fisica Cosmica, via P. Gobetti, 101, I40129 Bologna, Italy\and
    Universidad de Cantabria, Departamento de Ingenieria de Comunicaciones, av. de Los Castros s/n, 39005 Santander, Spain.\and
    Thales Alenia Space Italia S.p.A., S.S. Padana Superiore 290, 20090 Vimodrone, Milano, Italy\and
    SISSA/ISAS, Astrophysics Sector, Via Beirut 4, 34014 Trieste, Italy \and
    CESR, Centre d'Etude Spatiale des Rayonnements, 9, av du
Colonel Roche, BP 44346 31028 Toulouse Cedex 4, France\and
    Instituto de Fisica de Cantabria, CSIC, Universidad de Cantabria,
av. de los Castros s/n, 39005 Santander, Spain\and
    Jodrell Bank Centre for Astrophysics, Alan Turing Building, The University of Manchester, Manchester, M13 9PL, UK\and
    Herschel/Planck Project, Scientific Projects Dpt of ESA,
Keplerlaan 1, 2200 AG, Noordwijk, The Netherlands\and
    Istituto di Fisica del Plasma, CNR, via Cozzi 53, 20125 Milano, Italy\and
    ASI, Agenzia Spaziale Italiana, viale Liegi, 26, 00198 Roma, Italy \and
    Dipartimento di Fisica, Universit\`a degli Studi di Roma Tor Vergata, via della Ricerca Scientifica 1, 00133 Roma, Italy\and   
    INAF -- Osservatorio Astronomico di Padova, Vicolo dell'Osservatorio 5, 35122 Padova, Italy\and 
    INAF -- Osservatorio Astronomico di Trieste, via Tiepolo, 11, I-34143 Trieste, Italy
% ...
}

   \date{}

\abstract{In this paper we present the Low Frequency Instrument (LFI), designed and developed as part of the Planck space mission, the ESA program dedicated to precision imaging of the cosmic microwave background (CMB). Planck-LFI will observe the full sky in intensity and polarisation in three frequency bands centred at 30, 44 and 70 GHz, while higher frequencies (100-850 GHz) will be covered by the HFI instrument. The LFI is an array of microwave radiometers based on state-of-the-art Indium Phosphide cryogenic HEMT amplifiers implemented in a differential system using blackbody loads as reference signals. The front-end is cooled to 20K for optimal sensitivity and the reference loads are cooled to 4K to minimise low frequency noise. We provide an overview of the LFI, discuss the leading scientific requirements and describe the design solutions adopted for the various hardware subsystems. The main drivers of the radiometric, optical and thermal design are discussed, including the stringent requirements on sensitivity, stability, and rejection of systematic effects. Further details on the key instrument units and the results of ground calibration are provided in a set of companion papers.
}

       \keywords{
        Cosmology -– Cosmic Microwave Background –- Space Science -– Instrumentation
      }

%% file: 100_introduction.tex
Observations of the cosmic microwave background (CMB) have played a central role in the enormous progress of cosmology in the past few decades. Technological developments in both coherent radio receivers and bolometric detectors supported an uninterrupted chain of successful experiments, from the initial discovery \citep{penzias65} up to the present generation of precision measurements. Following COBE\footnote{ http://lambda.gsfc.nasa.gov/product/cobe/} and WMAP\footnote{http://map.gsfc.nasa.gov/}, the Planck\footnote{Planck (http://www.esa.int/Planck) is a project of the European
Space Agency - ESA - with instruments provided by two scientific Consortia funded by ESA member states (in particular the lead countries: France and Italy) with contributions from NASA (USA), and
telescope reflectors provided in a collaboration between ESA and a scientific Consortium led and funded by Denmark.} satellite, launched on 14 May 2009, is the next generation space mission dedicated to CMB observations. The Planck instruments are designed to extract all the cosmological information encoded in the CMB temperature anisotropies with an accuracy set by cosmic variance and astrophysical confusion limits, and to push polarisation measurements well beyond previously reached results. Planck will image the sky in nine frequency bands across the CMB blackbody peak, leading to a full-sky map of the CMB temperature fluctuations with signal-to-noise $>$10 and angular resolution $<10'$. The Planck instruments and observing strategy are devised to reach an unprecedented combination of angular resolution ($5'$ to $30'$), sky coverage (100\%), spectral coverage (27-900 GHz), sensitivity ($\Delta T/T \sim 2 \times 10^{-6}$), calibration accuracy ($\sim$0.5\%), and rejection of systematic effects \textbf ($\sim 1 \mu$K per pixel) \citep{2009_Tauber_Planck_Mission}. In addition, all Planck bands between 30 and 350 GHz will be sensitive to linear polarisation. 

The imaging power of Planck is sized to extract the temperature power spectrum with high precision over the entire angular range dominated by primordial fluctuations. This will lead to accurate estimates of  cosmological parameters that describe the geometry, dynamics and matter-energy content of the universe. The Planck polarisation measurements are expected to deliver complementary information on cosmological parameters and to provide a unique probe of the thermal history of the universe in the early phase of structure formation. Planck will also test the inflationary paradigm with unprecedented sensitivity through studies of non-Gaussianity and of B-mode polarisation as a signature of primordial gravitational waves \citep{planck_bluebook}.

The wide frequency range of Planck is required primarily to ensure accurate discrimination of foreground emissions from the cosmological signal. However, the nine Planck maps will also represent a rich data set for galactic and extragalactic astrophysics. Up to now, no single technology can reach the required performances in the entire Planck frequency range. For this reason two complementary instruments are integrated at the Planck focal plane exploiting state-of-the-art radiometric and bolometric detectors in their best windows of operation. The Low Frequency Instrument (LFI), described in this paper, covers the 27-77 GHz range with a radiometer array cooled to 20K. The High Frequency Instrument (HFI) will observe in six bands in the 90-900 GHz range with a bolometer array cooled to 0.1K \citep{2009_Lamarre_HFI}. The two instruments share the focal plane of a single telescope, a shielded off-axis dual reflector Gregorian system with 1.5$\times$1.9~m primary aperture \citep{2009_Tauber_Planck_Optics}.

The design of the Planck satellite and mission plan is largely driven by the extreme thermal requirements imposed by the instruments. The cold  payload enclosure ($<$50~K passive cooling) needs to be thermally decoupled from the warm ($\sim$300~K) service module while preserving high thermal stability. The optical design, orbit and scanning strategy are optimised to obtain the required effective angular resolution, rejection of stray-light and environmental stability. Planck will be injected into a Lissajous orbit around the Sun-Earth L2 point, at 1.5 million km from Earth. The scanning strategy assumes, to first order, the spacecraft spinning at 1~rpm with the spin axis aligned at 0$^\circ$ solar aspect angle (see \citet{2009_Tauber_Planck_Mission} for details). The typical angle between the detectors' line of sight and the spin axis is $\sim 85^\circ$. It will be possible to redirect the spin axis within a cone of $10^\circ$ around the spacecraft-sun axis. The baseline mission allows for 15 months of routine scientific operations in L2, a period in which the entire sky can be imaged twice by all detectors. However, in anticipation of a possible extension of the mission, spacecraft and instrument consumables allow an extension by a factor of two.

In this paper we present the design of the Planck-LFI and discuss its driving scientific requirements. We give an overview of the main subsystems, particularly those that are critical for scientific performance, while referring to a set of companion papers for more details. The LFI program as a whole, including data processing and programmatic issues, is described in \citet{2009_LFI_cal_M1}; the  calibration plan and ground calibration results are discussed by \citet{2009_LFI_cal_M3} and \citet{2009_LFI_cal_M4}. The LFI optical design is presented in \citet{2009_LFI_cal_M5}, while the expected polarisation performance is discussed in \citet{2009_LFI_polarisation_M6}. 

In section \ref{sec:frequency_range} we discuss the main scientific requirements of LFI. We will start from top-level guidelines such as frequency range, angular resolution and sensitivity, and then move to more detailed requirements that were derived for the chosen design  assuming a moderate level of extrapolation of the technology available at the time of the design completion.
In section \ref{sec:instrument_concept} we provide an overall description of LFI instrument configuration and discuss in detail the LFI  differential radiometers and associated components.  Section \ref{sec:lfi_configuration_subsystems} is a description of the instrument system and subsystems, including optical, radiometric and electronic units, while Sections \ref{sec:lfi_configuration_subsystems} to \ref{sec:optical_interfaces} describe the thermal, electrical and optical interfaces.

%% file: 200_scientific_requirements.tex
In this section we discuss the main scientific requirements for the LFI.
Note that here and throughout this paper we discuss instrument specifications, while measured on-ground performance are discussed in \citet{2009_LFI_cal_M3} and \citet{2009_LFI_cal_M4}. Measured values are generally in line with the design specifications, although noise levels are somewhat higher, particularly at 44 GHz. On the other hand, angular resolution at 70 GHz and stability at all frequencies surpass the requirement values. In the following sections we shall describe the design solutions implemented to meet such requirements.

%% file: 210_frequency_range.tex
The minimum of the combined diffuse emission of foregrounds relative to the CMB spectrum occurs at $\lambda\sim 4$~mm, i.e., roughly at the turning point between optimal performances of radiometric coherent receivers and bolometric detectors. Simulations carried out in the early design phases of Planck \citep{cobras_samba_redbook}
showed that a set of four logarithmically spaced bands in the 30-100 GHz range would provide a good spectral leverage to disentangle low frequency components while covering the window of minimum foregrounds for optimal CMB science.  The LFI is designed to cover the frequency range below the peak of the CMB spectrum using an array of differential radiometers \citep{Bersa_cobras_samba_1996,2000ApL&C..37..151M}. The initial Planck-LFI configuration \citep{Bers_Mand_LFI_2000} included four bands centred at 30, 44, 70 and 100~GHz, with the 100~GHz channel being covered by both LFI and HFI for scientific redundancy and systematics crosscheck. Budget and managerial difficulties, however, led to descoping of the LFI 100~GHz channel, which is now covered by HFI only. Nonetheless, the three LFI bands centred at 30, 44 and 70~GHz in combination with the six HFI bands provide Planck with a uniquely broad spectral coverage for robust separation of non-cosmological components. In addition, the LFI 70~GHz channel offers the cleanest view of the CMB for both temperature and polarisation anisotropy.

%% file: 220_angular_resolution_sensitivity.tex
Neglecting astrophysical foregrounds, calibration errors and systematic effects, and taking into account cosmic variance, the uncertainty in the parent distribution of the CMB power spectrum $C_\ell$ is given by \citep{knox95}:

\begin{equation}
    \frac{\delta C_\ell}{C_\ell} \simeq f_{\rm sky}^{-1/2} \sqrt {\frac {2}{2\ell+1}}\left[1+ \frac{A\sigma_{\rm pix}^{2}}{N_{\rm pix} C_\ell W_\ell^2}\right],
    \label{eq:delta_cl}
\end{equation}
where $\ell$ is the multipole index, $f_{\rm sky}$ is the fraction of CMB sky observed, $A$ is the surveyed area, $N_{\rm pix}$ is the number of pixels, $\sigma_{\rm pix}$ is the average noise per pixel at end of mission, and $W_\ell$ is the window function, which for LFI can be approximated by $W_\ell^2 =\exp \left[ - \ell ( \ell+1) \sigma_B^2\right]$ with $\sigma_B = \theta_{\rm FWHM} / \sqrt{8\ln2}=1.235\times10^{-4}\theta_{\rm FWHM}$ and $\theta_{\rm FWHM}$ is the full width half maximum of the beam, assumed  Gaussian, in arcmin. For a given mission lifetime, the noise per pixel in thermodynamic temperature is given by: 
\begin{equation}
    \sigma_{\rm pix}=\frac{\Delta T}{\sqrt{n_{\rm rad}\tau_{\rm pix}}},
    \label{eq:sigma_pix}
\end{equation}
where $\Delta T$ is the sensitivity of each radiometer of an array with $n_{\rm rad}$ elements, and 
\begin{equation}
    \tau_{\rm pix} = \frac{\tau_{\rm mission}}{N_{\rm pix}} = \frac{\tau_{\rm mission}}{4\pi/\theta^2_{\rm FWHM}}
    \label{eq:tau_pix} 
\end{equation}
is the integration time per resolution element in the sky.

%% file: 221_angular_resolution.tex
The basic scientific requirement for the Planck angular resolution is to provide approximately 10$'$ beams in the minimum foreground window, and to achieve up to 5$'$ in the highest frequency channels. This led to a telescope in the 1.5~m aperture class to ensure the desired resolution with an adequate rejection of straylight contamination \citep{mandolesi00,2002ExA....14....1V}. In general, a trade-off occurs between main beam resolution (half-power beam width, HPBW) and the illumination by the feeds of the edges of both the primary and sub-reflector (edge taper) which in turn drives the stray-light contamination effect. An edge taper $>$30~dB at an angle of 22$^\circ$ and an angular resolution of 14$'$ at 70~GHz were set as design specifications for LFI. Detailed calculations taking into account the location of the feeds in the focal plane and the telescope optical performance \citep{2009_LFI_cal_M5} showed that angular resolutions of $\sim 13'$ are achieved for the 70~GHz channels, while at lower frequencies we expect 24$'$--28$'$ at 44~GHz (depending on feed) and $\sim 33'$ at 30~GHz (see also Section \ref{sec:optical_interfaces}).  

%% file: 222_sensitivity.tex
To specify noise per frequency channel we adopted the general criterion of uniform sensitivity per equivalent pixel. In the early design phases, based on extrapolation of previous technological progress, we set a noise specification $\Delta T_{30}/T=3\times 10^{-6}$ (or $\Delta T_{30}=8\, \mu$K, thermodynamic temperature) for a reference pixel
$\Delta\theta_{30}\equiv 30'$ at all frequencies. We have also considered
``goal'' sensitivities of $\Delta T_{30}=6\,\mu$K per reference pixel, i.e., lower by 25\%. 

With an array of  $n_{\text{rad},\nu}$ radiometers at frequency  $\nu$ a sky pixel will be observed, on average, for an integration time

\begin{equation}
    \tau_{\text{tot},\nu}=\frac{n_{\text{rad},\nu}\theta_{\text{FWHM},\nu}^2}{4\pi}\tau_\text{mission}
\end{equation}

Assuming a 15 months survey, for $\Delta T_\text{30}=8\,\mu$K the sensitivity per pixel for a 1-s integration time is given by $\delta T_{1\text{sec}}\simeq 120\mu\text{K}\times
\sqrt{n_{\text{rad},\nu}}$.  We chose $n_{\text{rad},\nu}$ to compensate for the higher noise temperatures at higher frequencies while ensuring an  acceptable heat load in the Planck focal plane unit (see Section \ref{sec:active_cooling}), and allocate 4 radiometers at 30~GHz, 6 at 44~GHz and 12 at 70~GHz. 

As we will describe in detail in Sect.~\ref{sec:instrument_concept}, the LFI receivers are coupled in pairs to each feed horn ($n_\text{rad}=2\,n_\text{feeds}$) through an orthomode transducer. Thus the LFI design is such that all channels are inherently sensitive to polarisation. 
The sensitivity to $Q$ and $U$ Stokes parameters is lower than the sensitivity to total intensity $I$ by a factor  $\sqrt{2}$ since the number of channels per polarisation is only half as great. 
In order to optimise the LFI sensitivity to polarisation, the location and orientation of the LFI radiometers in the focal plan follows well defined constraints that are described in 
Sect.~\ref{sec:lfi_configuration_subsystems}. In Table~1 we summarise the main requirements for LFI sensitivity, angular resolution and the nominal LFI design characteristics.

\begin{figure}[tmb]
\begingroup
\newdimen\tblskip \tblskip=5pt
\vbox{{\bf Table~1.} LFI specifications for sensitivity and angular resolution.  $\Delta T_{30}$ indicates the noise per 30\arcm\ reference pixel.  Sensitivities per pixel are specified for a nominal mission survey time of 15~months.}
\nointerlineskip
\vskip -2mm
\footnotesize
\advance\baselineskip by 3pt 
\setbox\tablebox=\vbox{
   \newdimen\digitwidth 
   \setbox0=\hbox{\rm 0} 
   \digitwidth=\wd0 
   \catcode`*=\active 
   \def*{\kern\digitwidth}
   \newdimen\decimalwidth 
   \setbox0=\hbox{$.0$} 
   \decimalwidth=\wd0 
   \catcode`!=\active 
   \def!{\kern\decimalwidth}
\halign{\hbox to 1.4in{#\leaderfil}\tabskip=0.75em&
    \hfil#\hfil&
    \hfil#\hfil&
    \hfil#\hfil\tabskip=0pt\cr
\noalign{\doubleline}
\omit&30\,GHz&44\,GHz&70\,GHz\cr
\noalign{\vskip 3pt\hrule\vskip 5pt}
$\Delta T_{30}$ [$\mu$K]          &8&8&8\cr
$\Delta T_{30}/T$&$3\times10^{-6}$&$3\times10^{-6}$&$3\times10^{-6}$\cr
Angular resolution [\arcm]        &33&24&14\cr
$\Delta T/T$ per pixel            &$2.6\times10^{-6}$&$3.6\times10^{-6}$&        
        $6.2\times10^{-6}$\cr
$N_{\rm feeds}$                   &2&3&6\cr
$N_{\rm radiometers}$             &4&6&12\cr
$\delta T_{\rm 1\,s}^{\rm A}$ [$\mu$K\,s$^{1/2}$]$^{\rm a}$&234&278&365\cr
$\Delta\nu_{\rm eff}$ [GHz]       &6&8.8&14\cr
$T_{\rm sys}$ [K]$^{\rm b}$           &10.7&16.6&29.2\cr
\noalign{\vskip 5pt\hrule\vskip 3pt}}}
\enndtable
\tablenote a Antenna temperature.\par
\tablenote b Thermodynamic temperature.\par
\endgroup
\end{figure}

%% file: 230_sensitivity_budget.tex
For an array of coherent receivers, each with typical bandwidth $\Delta\nu$ and noise temperature $T_\text{sys}$, observing a sky antenna temperature  $T_{A,\text{Sky}}$, the average white noise per pixel (in antenna temperature) will be:

\begin{equation}
    \delta T_{\text{pix},A}=k_R\frac{T_\text{sys}+T_{A,\text{Sky}}}{\sqrt{\Delta \nu_\text{eff}\cdot \tau_\text{tot}}},
    \label{eq:delta_tpix}
\end{equation}
where  $k_R=\sqrt{2}$ for the LFI pseudo-correlation receivers. Therefore, for a required 
$\delta T_{\text{pix},A}$, the 1-second sensitivity (in antenna temperature) of each radiometer must be:

 \begin{equation}
    \delta T_{1\text{sec},A}(\mu\text{K}\sqrt{s})<\delta T_{\text{pix},A}\sqrt{n_\text{rad}\frac{\tau_\text{mission}}{N_\text{pix}}},
    \label{eq:delta_tpix_1sec}  
 \end{equation}
where $N_\text{pix}=4\pi/\theta_\text{FWHM}^2$ is the number of pixels.

%% file: 231_bandwidth_system_noise.tex
A first breakdown for contributions to LFI sensitivity is between system temperature and effective bandwidth. Each radiometer is characterised by a spectral response $g(\nu)$ which is determined by the overall spectral response of the system including amplifiers, waveguide components, filters, etc. We define the radiometer effective bandwidth as:

\begin{equation}
    \Delta\nu_\text{eff}=\frac{\left[\int^\infty_0 g(\nu)d\nu\right]^2}{\int^\infty_0 g^2(\nu)d\nu}
    \label{eq:delta_nu}
\end{equation}

In general, therefore, ripples in the band tend to narrow the ideal rectangular equivalent bandwidth. In practice, the effective bandwidth is limited by waveguide components, filters and in-band gain ripples. Pushing on available technology we assume for LFI a goal effective bandwidth of 20\% of the centre frequency. Equation \ref{eq:delta_tpix} then leads to requirements on $T_\text{sys}$ of $\sim$10~K at 30~GHz, and $\sim$30~K at 70~GHz (see Table~1).

%% file: 232_active_cooling.tex
These very ambitious noise temperatures can only be achieved with cryogenically cooled low noise amplifiers. Typically, the noise temperature of current state-of-the-art cryogenic transistor amplifiers exhibit a factor of 4--5 reduction going from 300~K to 100~K operating temperature, and another factor 2--2.5 from 100~K to 20~K. We implement active cooling to 20~K of the LFI front-end (including feeds, OMTs and first-stage amplification) to gain in sensitivity and to optimise the LFI-HFI thermo-mechanical coupling in the focal plane. 

Because the cooling power of the 20~K cooler (see Sect.~\ref{sec:thermal_interfaces}) is not compatible with the full radiometers operating at cryogenic temperature, each radiometer has been split into a 20~K front-end module and a 300~K back-end module, each carrying about half of the needed amplification ($\sim$70~dB overall). This solution also avoids the serious technical difficulty of introducing a detector operating in cryogenic conditions. A set of waveguides connect the front and back-end modules; these were designed to provide sufficient thermal decoupling between the cold and warm sections  of the instrument. Furthermore, low power dissipation components are required in the front-end. This is ensured by the new generation of cryogenic Indium Phosphide (InP) high electron mobility transistor (HEMT) devices, which yield world-record low noise performance with very low power dissipation.

%% file: 233_breakdown_allocations.tex
While the system noise temperature, $T_\text{sys}$, is dominated by the performance of front-end amplifiers, additional contribution comes from front-end losses and from back-end noise, which need to be minimised. For each LFI radiometer we can express the system temperature as follows:
\begin{equation}
    T_\text{sys}=T_\text{Feed+OMT}+T_\text{FE}+T_\text{WGs}+T_\text{BE},
    \label{eq:tsys_breakdown}
\end{equation}

where the terms on the right hand side represent the contributions from the feed\nobreakdash-horn/OMT, front\nobreakdash-end module, waveguides and back\nobreakdash-end module, respectively. These terms can be expressed as:
\begin{eqnarray}
    && T_\text{Feed+OMT} =\left(L_\text{Feed} L_\text{OMT} - 1\right)T_0\nonumber\\
    && T_\text{FE} = L_\text{Feed} L_\text{OMT} T_\text{FE}^{\text{noise}}\nonumber\\
    && T_\text{WGs}=\frac{L_\text{Feed} L_\text{OMT} \left(L_\text{WGs}-1\right)T_\text{eff}}{G_\text{FE}}\nonumber\\ 
    && T_\text{BE}=\frac{L_\text{Feed} L_\text{OMT} L_\text{WGs} T_\text{BE}^{\text{noise}}}{G_\text{FE}}\nonumber,
\end{eqnarray}
where $T_0$ is the physical temperature of the front-end; $T_\text{eff} \sim 200$K is an effective temperature of the waveguide whose exact value depends on the thermal design of the payload and radiometer interfaces; 
$T_\text{FE}^\text{noise}$ and $G_\text{FE}$ are the noise temperature and gain of the front-end module; $T_\text{BE}^\text{noise}$ is the back\nobreakdash-end module noise temperature; $L_\text{Feed}$, $L_\text{OMT}$ and $L_\text{WGs}$ are the ohmic losses from the feed, OMT and waveguides respectively, defined as
$L_\text{X}=10^{-L_\text{X,dB}/10}$ (for low loss components $L_\text{X} \gsim 1$ and $L_\text{X,dB} \lsim 0$).

In Table~2 we summarise the main LFI design allocations to the various elements contributing to the system temperature; these were established by taking into account state\nobreakdash-of\nobreakdash-the\nobreakdash-art technology.
Note that the contribution from front-end losses, $T_\text{Feed+OMT}$, is reduced to $\sim 15\%$ by cooling the feeds and OMTs to 20K and by using state-of-the-art low-loss waveguide components. Also, by requiring 30dB of gain in the radiometer front-end, noise temperatures for the back-end module of $\lsim 500$K (leading to $T_{BE} \lsim 0.5$K) can be acceptable which allows the use of standard GaAs HEMT technology for the ambient temperature amplification. More detailed design specifications for each component will be given in Section \ref{sec:lfi_configuration_subsystems} as we describe the instrument in further detail.

\begin{figure*}[tmb]
\begin{center}
\begingroup
\newdimen\tblskip \tblskip=5pt
\centerline{{\bf Table~2.} Sensitivity budget for LFI units.}
\nointerlineskip
\vskip -3mm
\footnotesize
\advance\baselineskip by 2pt 
\setbox\tablebox=\vbox{
   \newdimen\digitwidth 
   \setbox0=\hbox{\rm 0} 
   \digitwidth=\wd0 
   \catcode`*=\active 
   \def*{\kern\digitwidth}
   \newdimen\signwidth 
   \setbox0=\hbox{+} 
   \signwidth=\wd0 
   \catcode`!=\active 
   \def!{\kern\signwidth}
\halign{#\hfil\tabskip=2em&
\hbox to 2.0in{#\leaderfil}\tabskip=2em&
    \hfil#\hfil&
    \hfil#\hfil&
    \hfil#\hfil\tabskip=0pt\cr
\noalign{\doubleline}
\omit\hfil Symbol\hfil&\omit\hfil Name\hfil&30\,GHz&44\,GHz&70\,GHz\cr
\noalign{\vskip 3pt\hrule\vskip 5pt}
$L_{\rm Feed} L_{\rm OMT}$ [dB]&Feed + OMT insertion loss&$-$0.25&$-$0.25&$-$0.25\cr
$T_{\rm FE}$ [K]&FEM noise temperature&8.6&14.1&25.7\cr
$T_{\rm Feed}+T_{\rm OMT}+T_{\rm FE}$ [K]&&10.4&16.2&28.5\cr
$G_{\rm FE}$ [dB]&FEM gain&$>$30&$>$30&$>$30\cr
$L_{\rm WG}$ [dB]&Waveguide insertion loss&$-$2.5&$-$3&$-$5\cr
$T_{\rm BE}$ [K]&BEM noise temperature&350&350&450\cr
$T_{\rm WB}+T_{\rm BE}$ [K]&&0.3&0.4&0.7\cr
$T_{\rm sys}$ [K]&System temperature&10.7&16.6&29.2\cr
\noalign{\vskip 5pt\hrule\vskip 3pt}}}
\enndtable
\endgroup
\end{center}
\end{figure*}


%% file: 240_stability.tex
Considering perturbations to ideal radiometer stability, the minimum detectable temperature variation
of a coherent receiver is given by:
\begin{equation}
    \delta T(f)=k_R T_\text{sys}\sqrt{\frac{1}{\tau\cdot \Delta \nu_\text{eff}}+\left[\frac{\delta G_T(f)}{G_T}\right]^2},
    \label{eq:deltaT_with_gain}
\end{equation}
where $\delta G_T(f)/G_T$ represents the contribution from amplifier gain and noise temperature fluctuations at post-detection sampling frequency $f$. HEMT amplifiers are known to exhibit significant $1/f$ noise, caused by the presence of traps in the semiconductor \citep{jarosik96}, which would spoil the measurement if not suppressed. Amplifier fluctuations show a characteristic power spectral density $P(f)\propto 1/f^\alpha$ with $\alpha\approx 1$, so that the noise power spectral density is given by:

\begin{equation}
    P(f)\approx \sigma^2 \left[1+\left(\frac{f_k}{f}\right)^{\alpha}\right],
    \label{eq:delta_tf}
\end{equation}

where $\sigma^2$ represents the white noise limit and the knee-frequency, $f_k$ is the frequency at which the white noise and 1/$f$ components give equal contributions to the power spectrum (see \citet{2009_LFI_cal_R2} for a detailed discussion of the LFI noise properties). 

The 1/$f$ noise component not only degrades the sensitivity but also introduces spurious correlations in the time ordered data and sky maps. The reference frequency used to set a requirement on the knee frequency for LFI is the spacecraft spin frequency, 1 rpm, or 17 mHz. However, detailed analyses \citep{maino02a,Kei04} have shown that for the Planck scanning strategy a higher knee frequency ($f_k<50$~mHz) is acceptable as robust destriping and map making algorithms can be successfully applied to suppress the effects of low-frequency fluctuations. Because a total power HEMT receiver would have typical knee frequencies of 10 to 100~Hz, a very efficient differential design is needed for LFI in order to meet the 50~mHz requirement.

%% file: 250_systematics_requirements.tex
Throughout the design and development of LFI a key driver has been the minimisation and control of systematic effects, i.e., deviations from the signal that would be produced by an instrument with axially symmetric Gaussian beams, with ideal pointing and pure Gaussian white noise. These include optical effects (e.g., straylight, misalignment, beam distortions), instrument intrinsic effects (e.g., non-stationary and correlated noise features such as 1/$f$ noise, spikes, glitches, etc.), thermal effects (e.g., temperature fluctuations in the front-end or other instrument interfaces), and pointing errors. In particular, the LFI receiver (discussed in Section \ref{sec:instrument_concept}) was designed with the primary objective of minimising the impact of 1/f noise, thermal fluctuations and systematic effects due to non-ideal receiver components. 

The quantitative evaluation of various potential systematic effects required a complex iterative process involving design choices, knowledge and stability of the interfaces (with HFI and with the satellite), testing and modeling of the instrument behaviour, and simulations and simplified data analysis to evaluate the impact of each effect on the scientific output of the mission \citep{2004astro.ph..2528M}. Furthermore, dedicated analyses were required to evaluate the impact of instrument non-idealities on polarisation measurements \citep{2009_LFI_polarisation_M6}.

Limits on systematic effects impacting the effective angular resolution (beam ellipticity, alignment, pointing errors) were used, together with those coming from HFI, as input to the design of the Planck telescope and focal plane, as well as to set pointing requirements at system level. Regarding signal perturbations, for LFI we set an upper limit to the global impact of systematic effects of $<$3~$\mu$K per pixel at the end of the mission and after data processing. Starting from this cumulative limit, we defined a breakdown of contributions from various kinds of effects (Table~3), and then we worked out more detailed allocations for each contribution. This provided a useful guideline for the design, development and testing of the various LFI subsystems. For each type of systematic error we specify limits for three cases: a high frequency component, spin-synchronous fluctuations and periodic (non-spin-synchronous) fluctuations. High frequency contributions ($\gg$0.016~Hz) can be considered as random fluctuations and added in quadrature to the radiometers' white noise. As a goal, the overall noise increase due to random effects other than radiometer white noise should be less than 10\%. Spin synchronous (0.016~Hz) components are not damped by scanning redundancy, and impose the most stringent limits to systematic effects. Periodic fluctuations on time scales other than the satellite spin are damped with an efficiency that depends on the characteristic time scale of the effect (see \citet{mennella02} for quantitative analysis). For $1/f$ and thermal non-spin-synchronous fluctuations, affecting long time scales, we set the acceptable limits on systematic effects assuming that a consolidated destriping algorithm is applied to the data \citep{maino99, maino02a}.

\begin{figure}[tmb]
\begingroup
\newdimen\tblskip \tblskip=5pt
\vbox{\hsize=88mm{\bf Table~3.} Top-level systematic error budget (peak-to-peak values).  For $1/f$ and thermal periodic fluctuations the allocated limits are residuals after consolidated software removal techniques are applied to the data (numbers in parenthesis give the effect before removal).}
\nointerlineskip
\vskip -2mm
\footnotesize
\advance\baselineskip by 2pt 
\setbox\tablebox=\vbox{
   \newdimen\digitwidth 
   \setbox0=\hbox{\rm 0} 
   \digitwidth=\wd0 
   \catcode`*=\active 
   \def*{\kern\digitwidth}
   \newdimen\decimalwidth 
   \setbox0=\hbox{$.0$} 
   \decimalwidth=\wd0 
   \catcode`!=\active 
   \def!{\kern\decimalwidth}
\halign{\hbox to 1.4in{#\leaderfil}\tabskip=1.4em&
    \hfil#\hfil&
    \hfil#\hfil&
    \hfil#\hfil\tabskip=0pt\cr
\noalign{\doubleline}
\omit&Random&Spin synch&Periodic\cr
\noalign{\vskip 3pt\hrule\vskip 3pt}
\omit\hfil Source\hfil&$\delta T$ fraction&[$\mu$K]&[$\mu$K]\cr
\noalign{\vskip 3pt\hrule\vskip 5pt}
External straylight&$\ldots$&1&$\ldots$\cr
Internal straylight&0.045&1&0.9\cr
4\,K load          &0.025&1&0.6\,(11)\cr
Thermal fluctuations&0.03&0.8&1.1\,(11)\cr
Front end $1/f$    &0.25\,(0.34)&$\ldots$&$\ldots$\cr
Back end $1/f$     &0.35\,(0.48)&$\ldots$&$\ldots$\cr
DC electronics     &0.04&$\ldots$&$\ldots$\cr
Quantisation       &0.01&$\ldots$&$\ldots$\cr\cr
\noalign{\vskip 3pt}
\bf Total          &\bf 0.44&\bf1.9&\bf 1.5\cr
Noise increase&1.091\cr
\noalign{\vskip 5pt\hrule\vskip 3pt}}}
\enndtable
\endgroup
\end{figure}

High level allocations for signal perturbation effects are indicated in Table~3. The meaning of some of these contributions will become more clear as we provide a description of the design solutions adopted for the LFI instrument and its interfaces (Sections \ref{sec:lfi_configuration_subsystems} to \ref{sec:optical_interfaces}).

%% file: 300_instrument_concept.tex
The heart of the LFI instrument is an array of 22 differential receivers based on cryogenic high-electron-mobility transistor (HEMT) amplifiers. Cooling of the front-end is achieved by a closed-cycle hydrogen sorption cooler \citep{2009_LFI_cal_T2}, with a cooling power of about 1~W at 20~K, which also provides 18 K pre-cooling to the HFI.

\begin{figure}[h!]
    \begin{center}
        \includegraphics[width=7.5cm]{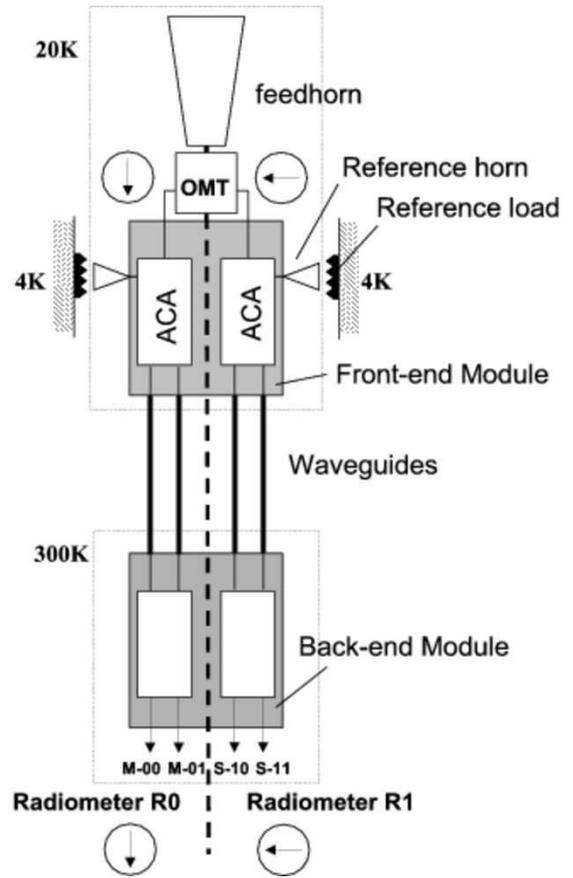}\\
\bigskip
        \includegraphics[width=8.5cm]{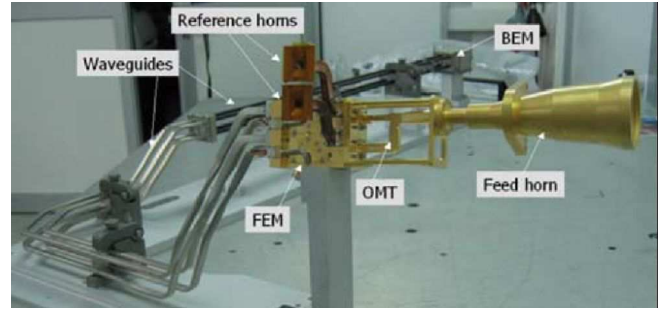}
    \end{center}
    \caption{Top: Schematic of a radiometer chain assembly (RCA). The LFI array has 11 RCAs, each comprising two radiometers carrying the two orthogonal polarisations. The RCA is constituted by a feed horn, an orthomode transducer (OMT), a front-end module (FEM) operated at~20 K, a set of four waveguides that connect FEM to the back-end module (BEM). The notations "0" and "1" for the two radiometers in the RCA denote the branches downstream the main and side arms of the OMT, respectively. Each amplifier chain assembly (ACA) comprises a cascaded amplifier and a phase switch. Bottom: picture of a 30~GHz RCA integrated before radiometer-level tests.}
    \label{fig:rca_schematic_and_picture}
\end{figure}

Radiation from the sky intercepted by the Planck telescope is coupled to 11  corrugated feed horns, each connected to a double-radiometer system, the so-called radiometer chain assembly (RCA, see Fig.~\ref{fig:rca_schematic_and_picture}). The complete LFI array, including 11 RCAs and 22 radiometers, is called the radiometer array assembly (RAA).

%% file: 310_rcas.tex
Downstream of each feedhorn, an orthomode transducer (OMT) separates the signal into two orthogonal polarisations with minimal losses and cross-talk. Two parallel independent radiometers are connected to the output ports of the OMT, thus preserving the polarisation information. Each radiometer pair is split into a front-end module (FEM) and a back-end module (BEM) to minimise power dissipation in the actively cooled front-end. A set of composite waveguides connect the FEM and the BEM.

The stringent stability requirements are obtained with a pseudo-correlation receiver in which the signal from the sky is continuously compared with the signal from a blackbody reference load. The loads, one for each radiometer, are cooled to approximately 4.5~K by a Stirling cooler that provides the second pre-cooling stage for the HFI bolometers. As we will show in detail in Section    \ref{sec:receiver_design}, each radiometer has two internal symmetric legs, so that each RCA comprises four waveguides connecting the FEM and the BEM, and four detector diodes. 

Each RCA is designated by a consecutive number (see Section \ref{sec:focal_plane_design}). In each RCA, the radiometer connected to the main-arm of the OMT is called R0, the one connected to the side-arm is called R1, as shown in Fig.~\ref{fig:rca_schematic_and_picture}. 
The two detectors in radiometer R0 are named (M-00,M-01), while those in radiometer R1 are named (S-10,S-11).

%% file: 320_raa.tex
\begin{figure}[h!]
\begin{center}
    \includegraphics[width=8.5cm]{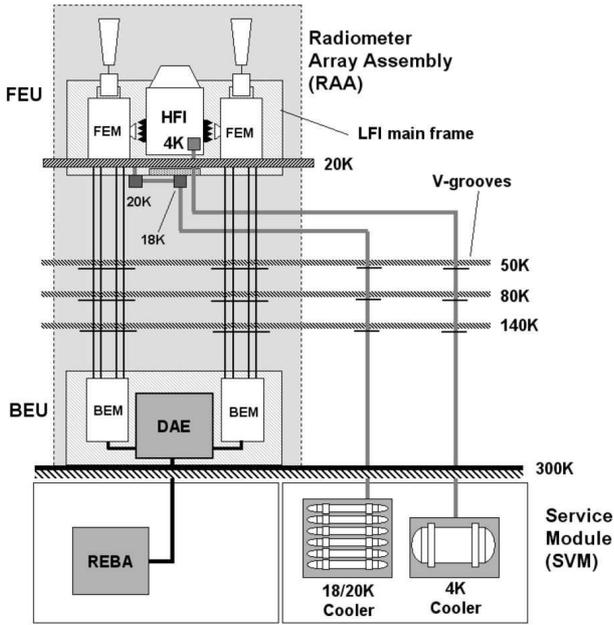}
\end{center}
    \caption{Schematic of the LFI system displaying the main thermal interfaces with the V-grooves and connections with the 20~K and 4~K coolers. Two RCAs only are shown in this scheme. The Radiometer Array Assembly (RAA) is represented by the shaded area and comprises the front-end unit (FEU) and back-end unit (BEU). The entire LFI RAA includes 11 RCAs, with 11 feeds, 22 radiometers, and 44 detectors.}
    \label{fig:raa_schematic}
\end{figure}

A schematic of the Radiometer Array Assembly (RAA), is shown in Fig.~\ref{fig:raa_schematic}. Each RCA has been integrated and tested separately, and then mounted on the RAA without de-integration to ensure stability of the radiometer characteristics after calibration at RCA level \citep{2009_LFI_cal_M4}.

A ''mainframe" supports the LFI 20~K front-end (with feeds, OMTs and FEMs) and interfaces the HFI 4~K front-end box in the central portion of the focal plane. The HFI 4~K box is linked to the 20~K LFI
mainframe with insulating struts, and provides the thermal and mechanical interface to the LFI reference loads. Fourty-four waveguides connect the LFI front-end unit (FEU) to the back-end unit (BEU) which
is mounted on the top panel of the Planck service module (SVM) and is maintained at a temperature of $\sim 300$~K. The BEU comprises the eleven BEMs and the data acquisition electronics (DAE) unit. After on-board processing, provided by the Radiometer Box Electronics Assembly (REBA), the compressed signal is down-linked to the ground station together with housekeeping data.

A major design driver has been to ensure acceptably low conductive and radiative parasitic thermal loads on the 20K stage, particularly those introduced by the waveguides and cryo-harness. As we will discuss below, sophisticated design solutions were implemented for these units. In addition, three thermal sinks were used to largely reduce the parasitic loads in the 20K stage; these are the three conical shields (V-grooves) introduced in the Planck payload module to thermally isolate the cold telescope enclosure from the SVM at $\sim$~300~K \citep{2009_Tauber_Planck_Mission}. The V-grooves also provide multiple precool temperatures to all of the Planck coolers, as well as intercepting parasitics from the cooler piping and HFI equipment. The three V-grooves  are expected to reach in-flight temperatures of approximately 170~K, 100~K and 50~K, respectively. 

\begin{figure}[h!]
    \begin{center}
        \includegraphics[width=8.5cm]{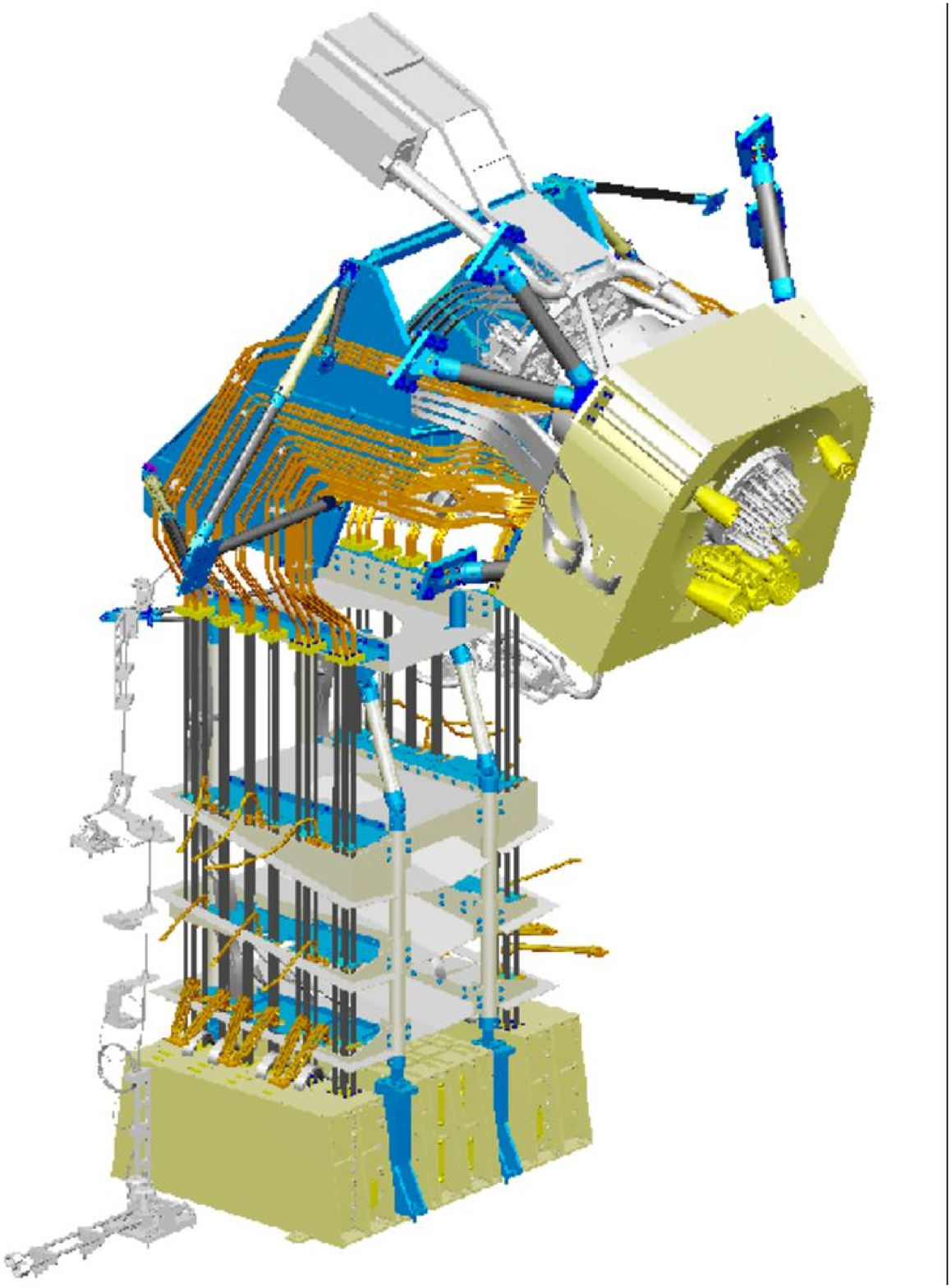}\\
        \includegraphics[width=8.5cm]{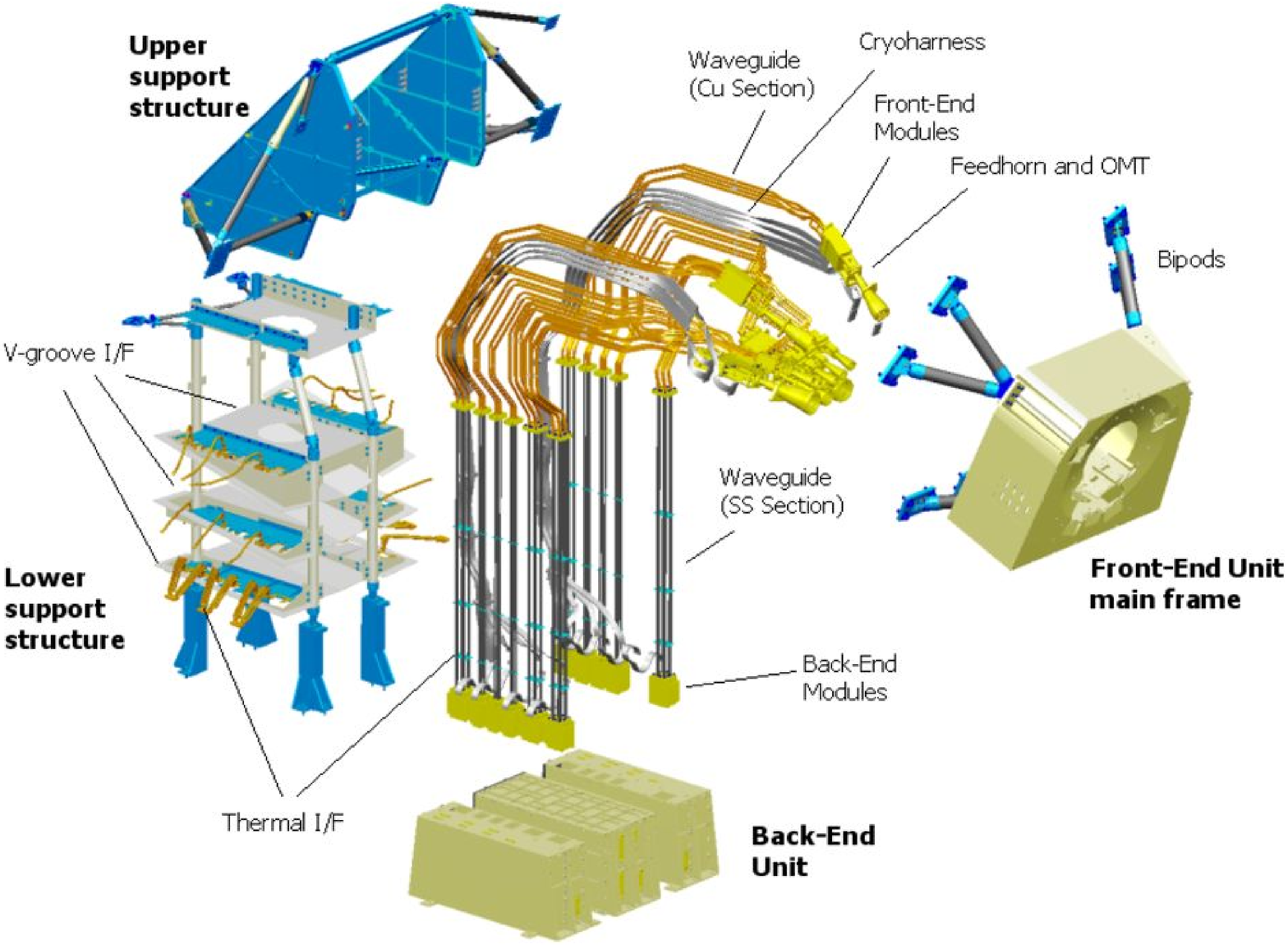}\\
    \end{center}
    \caption{LFI Radiometer Array Assembly (RAA). Top: drawing of the integrated instrument showing the focal plane unit, waveguide bundle and back-end unit. The elements that are not part of LFI hardware (HFI front-end, cooler pipes, thermal shields are shown in light gray. Bottom: More details are visible in the exploded view, as indicated in the labels.}
    \label{fig:raa_3d}
\end{figure}

\begin{figure}[h!]
    \begin{center}
        \includegraphics[width=7.5cm]{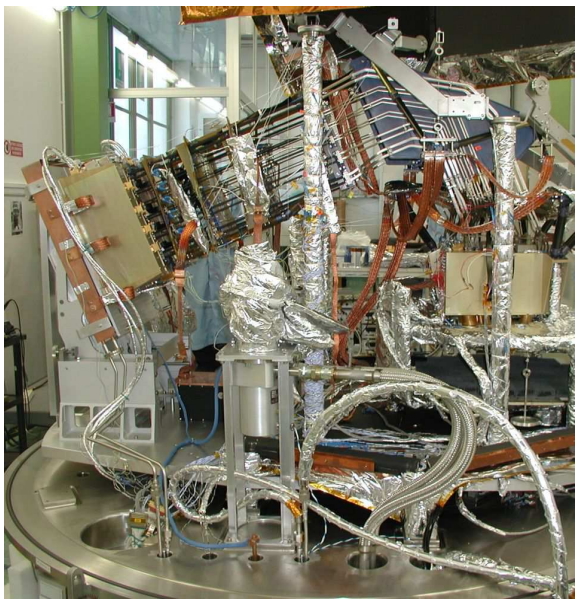}
    \end{center}

    \caption{The LFI instrument in the configuration for instrument level test cryogenic campaign.}
    \label{fig:raa_cryofacility}
\end{figure}

\begin{figure}[h!]
    \begin{center}
        \includegraphics[width=7.5cm]{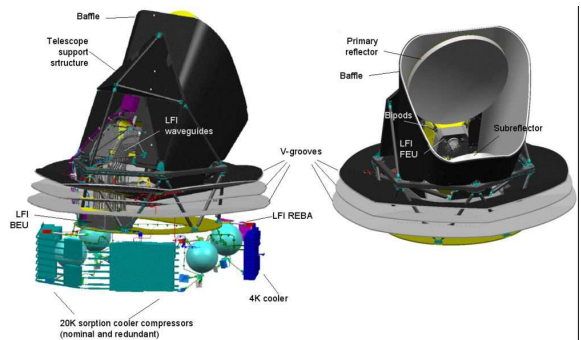}\\
        \includegraphics[width=7cm]{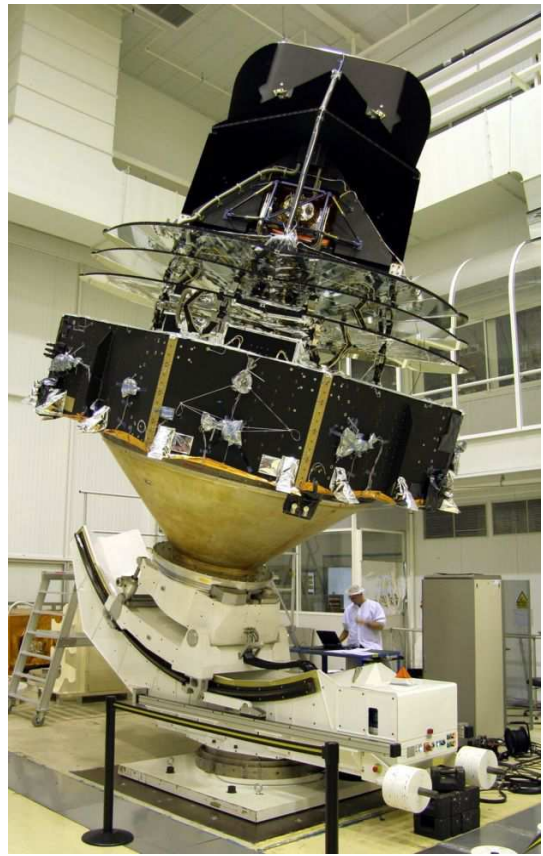}
    \end{center}
    \caption{Top: schematic of the Planck satellite showing the main interfaces with the LFI RAA on the spacecraft. 
Bottom: back view of Planck showing the RAA integrated on the PPLM. The LFI Back-end unit is the box below the lowest V-groove and resting on the top panel of the SVM.}
    \label{fig:lfi_on_planck}
\end{figure}

The FEU is aligned in the focal plane of the telescope and supported by a set of three thermally insulating bipods attached to the telescope structure. The back-end unit is fixed on top of the Planck service module, below the lower V-groove. In Fig.~\ref{fig:raa_3d} we show a detailed drawing of the RAA, including an exploded view showing its main sub-assemblies and units. After integration the RAA was first tested in a dedicated cryo-facility \citep{2009_LFI_cal_T2} for instrument level tests (Fig.~\ref{fig:raa_cryofacility}), and then inserted in the payload module after integration of the HFI 4~K box.  Fig.~\ref{fig:lfi_on_planck} shows the LFI within the Planck satellite.

%% file: 330_receiver_design.tex
The LFI receivers are based on an additive correlation concept, or pseudo-correlation, analogous to schemes used in previous applications in early works \citep{Blum1959} as well as in recent CMB experiments \citep{1996ApJ...458..407S,jarosik03}. The LFI design introduces new features that optimise stability and immunity to systematics within the constraints imposed by cryogenic operation and by integration into a complex payload such as Planck. The FEM contains the most sensitive part of the receiver, where the pseudo-correlation scheme is implemented, while the BEM provides further RF amplification and detection. 
    
\begin{figure}[h!]
    \begin{center}
        \includegraphics[width=9cm]{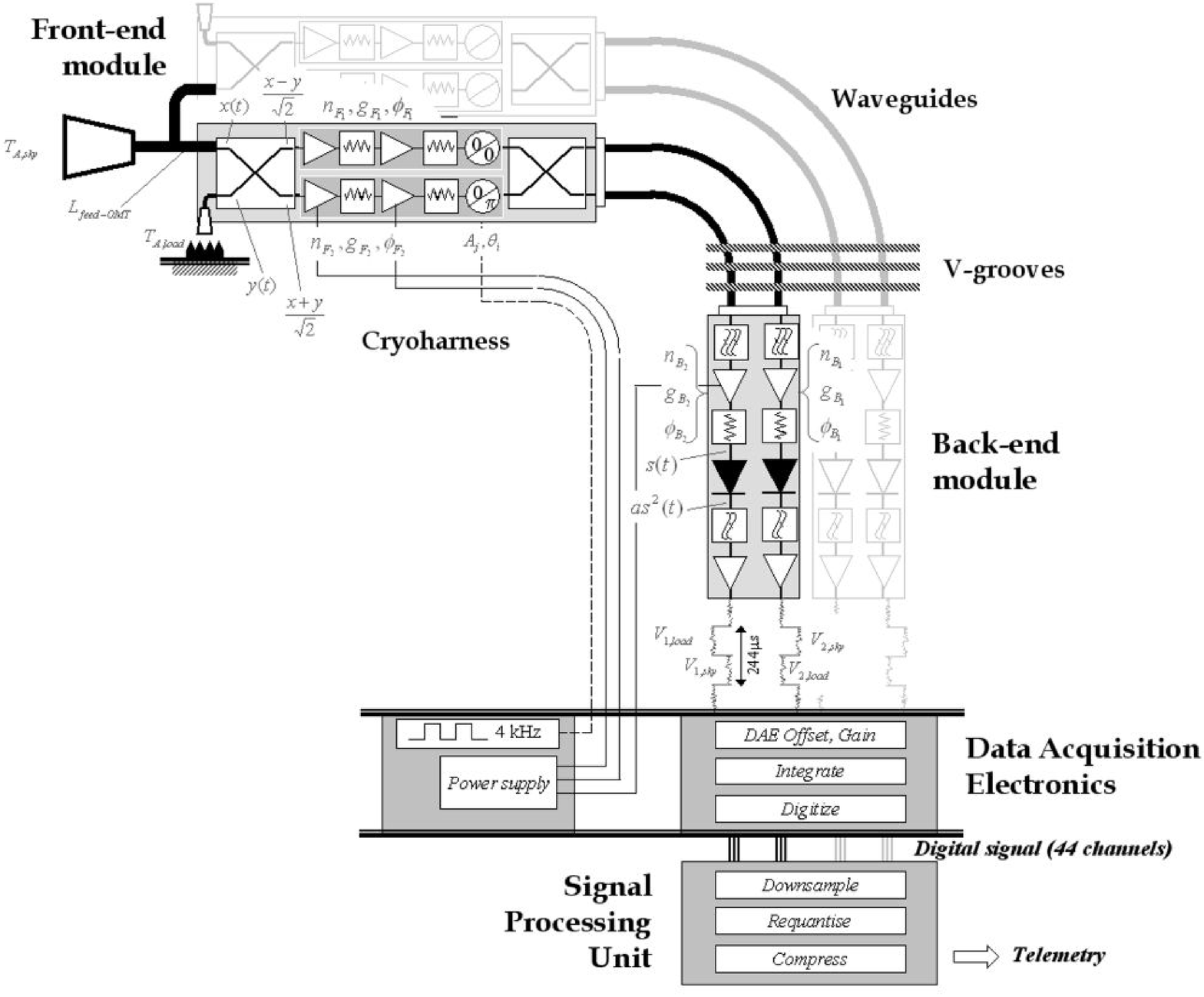}
    \end{center}
    \caption{LFI receiver scheme, shown in the layout of a Radiometer Chain Assembly (RCA). Some details in the receiver components (e.g., attenuators, filters, etc) differ slightly for the different frequency bands.}
    \label{fig:rca_schematics_big}
\end{figure}

In each radiometer (Fig.~\ref{fig:rca_schematics_big}), after the OMT, the voltages of the signal from the sky horn, $x(t)$, and from the reference load, $y(t)$, are coupled to a 180$^\circ$ hybrid which yields the mixed signals $(x+y)/\sqrt{2}$ and $(x-y)/\sqrt{2}$ at its two output ports. These signals are then amplified by the cryogenic low noise amplifiers (LNAs) characterised by noise voltage, gain and phase $n_{F_1}$, $g_{F_1}$, $\phi_{F_1}$ and $n_{F_2}$, $g_{F_2}$, $\phi_{F_2}$. One of the two signals then runs through a switch that shifts the phase between 0 and 180$^\circ$ at a frequency of 4096~Hz. A second phase switch is mounted for symmetry and redundancy on the other radiometer leg, but it does not introduce any switching phase shift. The signals are then recombined by a second 180$^\circ$ hybrid coupler, thus producing an output which is a sequence of signals proportional to $x(t)$ and $y(t)$ alternating at twice the phase switch frequency.

In the back-end modules (Fig.~\ref{fig:rca_schematics_big}), the RF signals are further amplified in the two legs of the radiometers by room temperature amplifiers characterised by noise voltage, gain and
phase $n_{B_1}$, $g_{B_1}$, $\phi_{B_1}$ and $n_{B_2}$, $g_{B_2}$, $\phi_{B_2}$. The signals are filtered and then detected by square-law detector diodes. A DC amplifier then boosts the signal output which is connected to the data acquisition electronics. The sky and reference load DC signals are integrated, digitised and then transmitted to the ground as two separated streams of sky and reference load data.
    
The sky and reference load signals recombined after the second hybrid in the FEM have highly correlated 1/$f$ fluctuations. This is because, in each radiometer leg, both the sky and the reference signals undergo the same instantaneous fluctuations due to LNAs intrinsic instability. Furthermore, the fast modulation drastically reduces the impact of 1/$f$ fluctuations coming from the back-end amplifiers and detector diodes, since the switch rate $\sim$4~kHz is much higher than the 1/$f$ knee frequency of the BEM components. By
taking the difference between the DC output voltages $V_\text{sky}$ and $V_\text{load}$, therefore, the 1/$f$ noise is highly reduced. 

Differently from the WMAP receivers, the LFI phase switches and second hybrids have been placed in the front end. This allows full modularity of the FEMs, BEMs and waveguides, which in turns simplifies the integration and test procedure. Furthermore, this design does not require that the phase be preserved in the waveguides, a major advantage given the complex routing imposed by the LFI-HFI integration and the potentially significant thermo-elastic effects from the cryogenic interfaces in the Planck payload.

In principle, for a null differential output corresponding to a perfectly balanced system, fluctuations would be fully suppressed in the differenced data. In practice, for LFI, a residual offset will be necessarily present due to input asymmetry between the sky arm ($\sim 2.7$~K from the sky, plus $\sim$0.4~K from the
reflectors) and the reference load arm (with physical temperature $\sim$4.5K), plus a small contribution from inherent radiometer asymmetry. To compensate for this effect, a gain modulation factor $r$ is introduced in software to null the output by taking the difference
$\bar p=V_\text{sky}-r\, V_\text{load}\approx 0$. In the next section we will discuss the
signal model in more detail.

%% file: 331_signal_model.tex
If $x(t)$ and $y(t)$ are the input voltages at each component, then the transfer functions for the hybrids, the front-end
amplifiers, the phase switches and the back-end amplifiers can be written, respectively, as:
\begin{eqnarray}
    &&f_\text{hybrid}:\left\{x,y\right\}\rightarrow     
        \left\{\frac{x+y}{\sqrt{2}},\frac{x-y}{\sqrt{2}}\right\}\\
    &&f_\text{amp}^\text{FE}:\left\{x,y\right\}\rightarrow
        \left\{g_{F_1}(x+n_{F_1})e^{i\,\phi_{F_1}},
               g_{F_2}(y+n_{F_2})e^{i\,\phi_{F_2}}\right\}\nonumber\\
    &&f_\text{sw}:\left\{x,y\right\}\rightarrow
        \left\{x,y\sqrt{A_j}e^{i\, \theta_j}\right\},j=1,2\nonumber\\
    &&f_\text{amp}^\text{BE}:\left\{x,y\right\}\rightarrow
        \left\{ g_{B_1}(x+n_{B_1})e^{i\, \phi_{B_1}},
                g_{B_2}(y+n_{B_2})e^{i\, \phi_{B_2}}\right\},\nonumber
    \label{eq:transfer_functions}
\end{eqnarray}
where  $\theta_1$ and $\theta_2$ are the phase shifts in the two switch states (nominally, $\theta_1=0$ and $\theta_2=180^\circ$),  $n_F$ and $n_B$ represent the white noise of the front-end and back-end amplifiers, and $A_1$ and $A_2$ represent the fraction of the signal amplitude that is transmitted after the phase switch in the two states (for a lossless switch $A_1=A_2=1$). Based on these transfer functions and on the topology of the LFI receiver discussed above, we developed a detailed analytical description of the receiver and evaluated its susceptibility to systematic effects by studying the impact of deviation from ideal radiometer behaviour on the differenced output \citep{seiffert02, mennella02}. In addition to the analytical treatment, a numerical model of the RCA signals has been developed \citep{2009_LFI_cal_R5}.

For small phase mismatches and assuming negligible phase switch imbalance, the power output of the differenced signal after applying the gain modulation factor is given by:

\begin{eqnarray}
    \Delta p &=& a\, k\, \beta \left[\tilde T_{A,\text{sky}}(G-r\,I)-r \tilde T_{A,\text{load}}\left(G-\frac{1}{r}I\right)\right.+\nonumber\\
    &+&\left.(1-r)T_\text{sys}\right].
    \label{eq:delta_p}
\end{eqnarray}
In Eq.~(\ref{eq:delta_p}) $a$ is the proportionality constant of the square law detector diode, $G$ and $I$ are the effective power gain and isolation of the  system:
\begin{eqnarray}
    G\simeq \frac{1}{4}g_B^2(g_{F_1}^2+g_{F_2}^2+2g_{F_1}g_{F_2})\nonumber\\
    I\simeq \frac{1}{4}g_B^2(g_{F_1}^2+g_{F_2}^2-2g_{F_1}g_{F_2}),\nonumber\\
    \label{eq:gain_and_isolation}
\end{eqnarray}
where $g_B$ is the voltage gain of the BEM in the considered channel. In Eq.~(\ref{eq:delta_p}) the temperature terms:
\begin{eqnarray}
    &&\tilde T_\text{sky}=
        \frac{T_\text{sky}}{L_\text{feed}L_\text{OMT}}+\left(1-\frac{1}{L_\text{feed}L_\text{OMT}}\right)
        T_\text{phys}\nonumber\\
    &&\tilde T_\text{load}=
        \frac{T_\text{load}}{L_\text{4K}}+\left(1-\frac{1}{L_\text{4K}}\right)T_\text{phys},
    \label{eq:tilde_temperatures}
\end{eqnarray}
represent the sky and reference load signals at the input of the first hybrid, where $L_\text{4K}$ is the insertion loss of the reference horn, and $T_\text{phys}\simeq 20$~K is the front-end physical temperature.

\begin{figure*}
    \begin{center}
        \includegraphics[width=15cm]{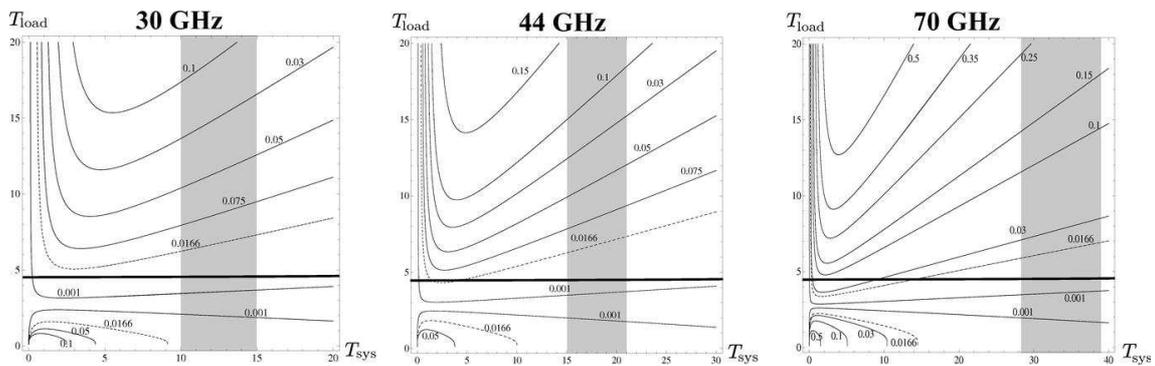}
    \end{center}
    \caption{Curves of equal $f_k$ (in Hz) on the plane $T_{load}$ (K, thermodynamic temperature), $T_{sys}$ assuming thermodynamic sky temperature of 2.7~K. Each panel refers to a different frequency channel. The dashed contour refers to values for which the knee frequency is equal to the spin frequency ($f_{\rm spin} = 0.0166$~Hz). The graphs also show the range of typical LFI noise temperature values (grey area) and the nominal reference load temperature (4~K - double horizontal line).}
    \label{fig:one_over_f_plots}
\end{figure*}

%% file: 332_knee_frequency_and_gmf.tex
For a radiometer with good isolation ($>$13~dB), as expected in a well matched system, it follows from Eq.~(\ref{eq:delta_p}) that the power output is nulled for

\begin{equation}
    r=\frac{\tilde T_\text{sky}+T_\text{noise}}{\tilde T_\text{load}+T_\text{noise}}.
    \label{eq:gmf}
\end{equation}

In this case the gain fluctuations are fully suppressed and the radiometer is sensitive only to the 1/$f$ noise caused by noise temperature fluctuations, which represents only a small fraction of the amplifiers instability. For an optimal choice of the gain modulation factor, the resulting knee frequency is given by:

\begin{equation}
    f_k\simeq \Delta\nu\left(\frac{A(1-r)T_\text{sys}}{T_\text{sky}+T_\text{sys}}\right)^2\propto (1-r)^2.
    \label{eq:optimal_fk}
\end{equation}

Thus, in principle, for small input offsets $\tilde T_\text{sky}\simeq \tilde T_\text{load}$ very small knee frequencies can be obtained. Fig.~\ref{fig:one_over_f_plots} displays expected knee frequencies for parameters typical of the LFI channels as a function of noise temperature and reference load temperature, assuming ideal gain and phase match. Note that for $T_\text{load}\approx 4$~K we expect $f_k$ an order of magnitude lower than for $T_\text{load}\approx 20$~K. This was the driver for implementing reference loads at 4K at the cost of some complexity in the thermo-mechanical interfaces in the focal plane. It can also be shown \citep{2001_mennella_back_end_fluctuations} that the same value of $r$ that minimises the radiometer sensitivity to 1/$f$ noise is also effective in reducing the susceptibility to other systematic effects such as back-end temperature variations.

\begin{figure}[h!]
    \begin{center}
        \includegraphics[width=7.5cm]{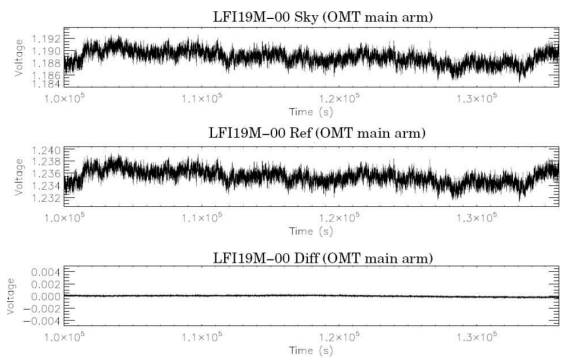}
    \end{center}
    \caption{Example of uncalibrated data stream from one of the 44 LFI detectors (LFI19M-00, at 70~GHz) recorded during instrument level tests. The upper and middle panels show the data for the sky and reference load inputs, while the lower panel shows differenced data stream with optimal gain modulation factor.}
    \label{fig:tods}
\end{figure}

It is essential that the gain modulation factor $r$ be calculated with sufficient precision to reach the required stability. Simulations and testing show that the needed accuracy ranges from $\pm 1\%$ (30~GHz) to $\pm 0.5\%$ (70~GHz). This accuracy can be obtained with different methods \citep{mennella03}, the simplest being to evaluate the ratio of the total power output voltages averaged over a suitable time interval, $r\simeq \bar V_\text{sky}/\bar V_\text{load}$. In Fig.~\ref{fig:tods} we show, as an example, the data streams from one of the 44 LFI detectors with the two total power signals and differenced data. The LFI telemetry allocations ensure that the total power data from both the sky and reference load samples will be downloaded, so calculation of $r$ and differencing will be performed on the ground. 

Further suppression of common fluctuation modes, typically of thermal or electrical origin, is obtained by taking the noise-weighted average of the two detectors associated to each radiometer (see Appendix E in \citet{2009_LFI_cal_M3}) as well as in the differencing of the main and side arm radiometers signals when analysing data for polarisation 
\citep{2009_LFI_polarisation_M6}.

%% file: 333_noise_temperature.tex
The LFI radiometer sensitivity is essentially independent of the temperature of the reference loads. From Equations~(\ref{eq:delta_tpix}) and (\ref{eq:delta_p}) it follows that, to first order, the radiometer sensitivity is:
\begin{equation}
    \delta T =\sqrt{\frac{2}{\tau \cdot \Delta \nu}} (T_\text{sys}+T_\text{sky})\sqrt{1+\eta_L},
    \label{eq:delta_Trms}
\end{equation}

where 

\begin{equation}
    \eta_L=\frac{(T_{N,B}/G_F)^2}{(T_\text{sys}+T_\text{sky})(T_\text{sys}+T_\text{load})},
    \label{eq:eta_l}
\end{equation}
and  $T_{N,B}$ is the noise temperature of the back-end, $G_F$ is the gain of the front end. For parameter values typical of LFI we have $\eta_L< 10^{-3}$, so that the dependence of
the noise temperature on $T_\text{load}$ is extremely weak. The advantage of cooling the reference load to 4~K, therefore, rests solely on better suppression of systematics, not on
sensitivity.

%% file: 400_configuration_subsystems.tex
\begin{figure*}
    \begin{center}
        \includegraphics[width=20cm]{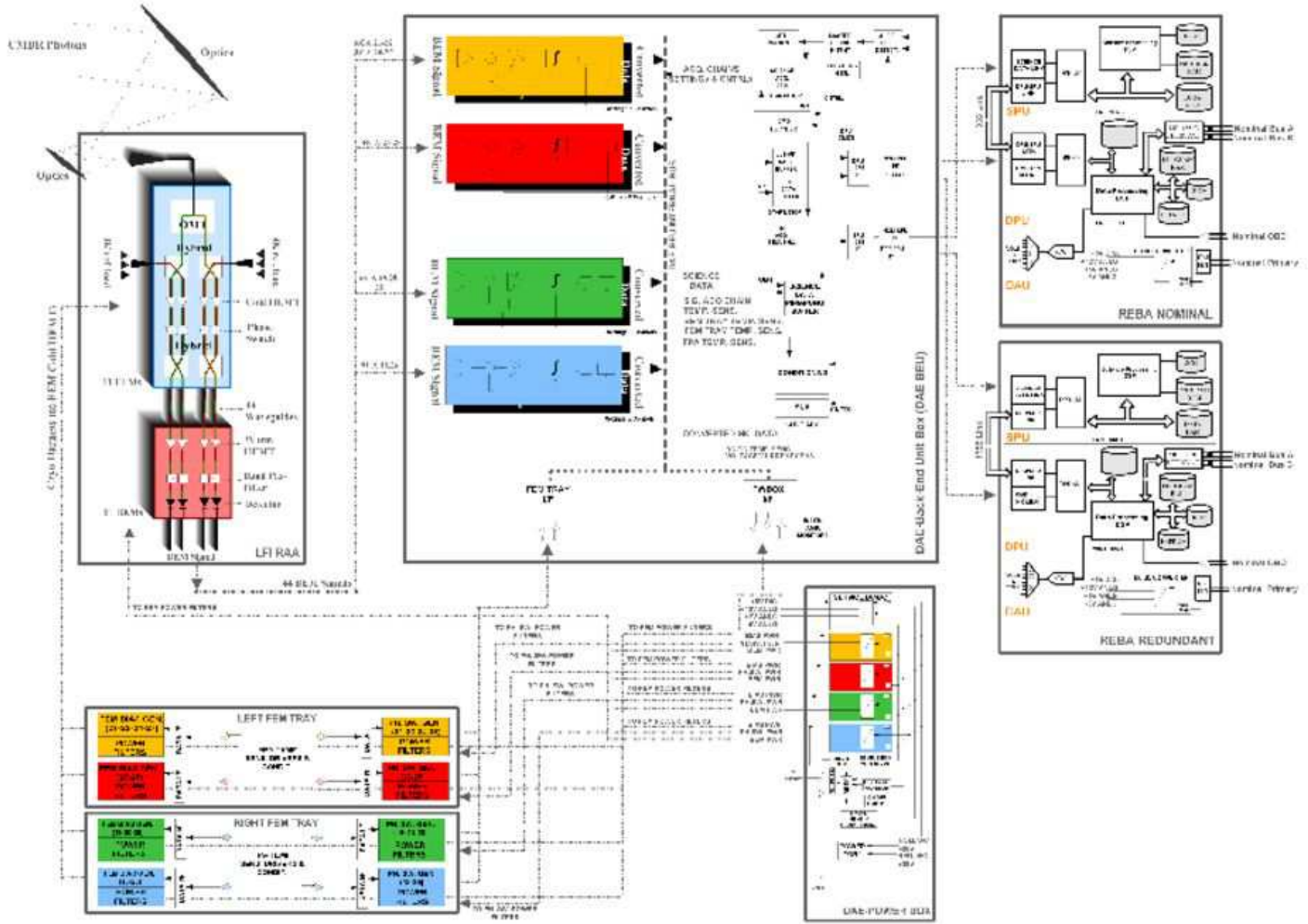}
    \end{center}
    \caption{Schematic of the LFI showing interconnections and details of the DAE and REBA main functions and units.}
    \label{fig:lfi_block_diagram}
\end{figure*}

The overall LFI system is shown schematically in Fig.~\ref{fig:lfi_block_diagram}. In this section we provide an overview of the instrument units and main subsystems. Further details on the design, development and testing of the most critical components are given in companion papers that will be referenced below.

%% file: 411_focal_plane_design.tex
The disposition of the LFI feeds in the focal plane is driven by optimisation of angular resolution and of recovery of polarisation information. In addition, requirements on proper sampling of the sky and rejection of crosstalk effects need to be met.

\begin{figure}[h!]
    \begin{center}
        \includegraphics[width=7.5cm]{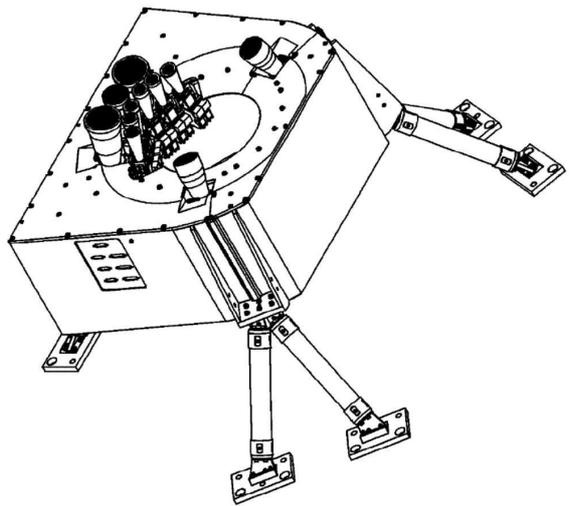}\\
\bigskip
        \includegraphics[width=7.5cm]{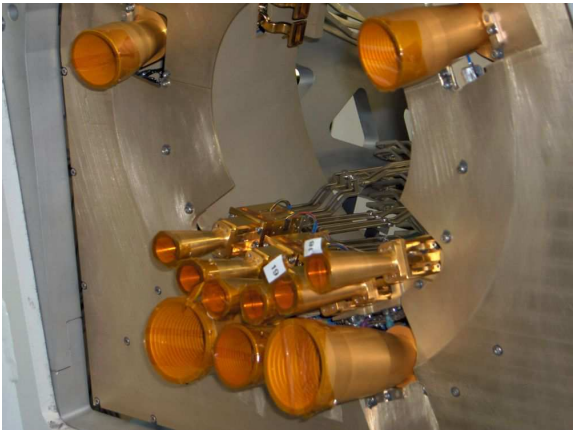}
    \end{center}
    \caption{
Arrangement of the LFI feeds in the focal plane. 
Top: 
mechanical drawing of the main frame and focal plane elements. Shown are the bipods connecting the FPU to the telescope structure. 
Bottom: 
picture of the LFI flight model focal plane.}
    \label{fig:lfi_focal_plane}
\end{figure}

The central portion of the Planck focal plane is occupied by the HFI front end, as higher frequency channels are more susceptible to optical aberration. The LFI feeds are located as close as possible to the focal plane centre compatible with mechanical interfaces with HFI and 4~K reference loads (Fig.~\ref{fig:lfi_focal_plane}). Miniaturised designs for the FEMs and OMT are implemented to allow optimal use of the focal area. The 70~GHz feeds, most critical for cosmological science, are placed in the best location for angular resolution and low beam distortion required for this frequency \citep{2009_LFI_cal_M5}.

\begin{figure}[h!]
    \begin{center}
        \includegraphics[width=7.5cm]{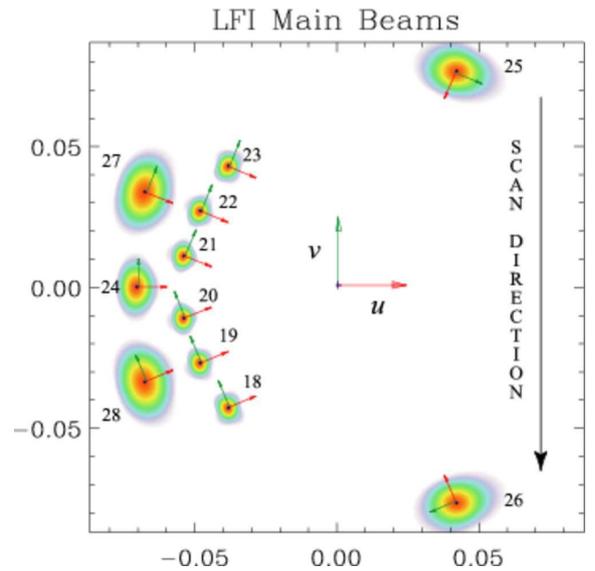}
    \end{center}
    \caption{
Footprint of the LFI main beams on the sky and polarisation angles as seen by an observer looking towards the satellite along its optical axis.
The units are $u$-$v$ coordinates defined as $u=\sin\theta_{bf}\cos\phi_{bf}$ and 
$v=\sin\theta_{bf}\sin\phi_{bf}$, where $\theta_{bf}$ and $\phi_{bf}$ are referred to each main beam frame (see \citet{2009_LFI_cal_M5}. The angular region covered by the plot is approximately $10^\circ \times 10^\circ$. Labels from 18 to 23 refer to 70 GHz horns, from 24 to 26 refer to 44 GHz horns, and 27 and 28 refer to 30GHz horns. The scan direction, orthogonal to the focal plane symmetry axis, is indicated by an arrow. All the relative angles in pairs of feeds aligned along the scan direction are shifted by 45$^\circ$, with the exception of RCA24.}
    \label{fig:lfi_beams}
\end{figure}

A key criterion for the feed arrangement is that the $E$ and $H$ planes as projected in the sky will allow optimal discrimination of the Stokes $Q$ and $U$ parameters. The polarisation information is obtained by differencing the signal measured by the main-arm (R0) and side-arm (R1)  radiometers in each RCA, which are set to 90$^\circ$ angle by the OMT. An optimal extraction of $Q$ and $U$ is achieved if subsets of channels are oriented in such a way that the linear polarisation directions are evenly sampled 
\citep{2009_LFI_polarisation_M6}. We achieve this by arranging pairs of feeds with their $E$ planes oriented at 45$^\circ$ to each other (Fig.~\ref{fig:lfi_beams}). 
With the exception of one 44~GHz feed, all the $E$ planes are oriented at $\pm$22.5$^\circ$ relative to the scan direction. We also align pairs of feeds such that their projected lines of sight will follow each other in the Planck scanning strategy.

The spin axis of Planck will be shifted by $=2'$ every  $\sim$ 45 min \citep{2009_Tauber_Planck_Mission}. Therefore, even for the highest LFI angular resolution,  $\theta_\text{FWHM}\simeq 13'$ at 70~GHz, the sky is well sampled.

%
%

%% file: 412_lfi_main_frame.tex
The LFI mainframe provides thermo-mechanical support to the LFI radiometer front-end, but it also supports the HFI and interfaces to the cold end (nominal and redundant) of the 20~K sorption cooler. In fact, some of the key requirements on the LFI mainframe (stiffness, thermal isolation, optical alignment) are driven by the HFI rather than the LFI instrument. The mainframe is built of Aluminium alloy 6061-T6 and it is  dismountable in three sub-units to facilitate integration of the complete RCAs and of the HFI front-end.

The interface between the FPU and the 50~K payload module structure must ensure thermal isolation as well as compliance with eigenfrequencies from launch loads. A trade-off was made to identify the proper material properties, fibre orientations, and strut inclinations. The chosen configuration was a set of three 225-mm-long CFRP T300 bipods, inclined at 50$^\circ$. The LFI-HFI interface is provided by a structural ring connected to the HFI by six insulating struts locked to the LFI main frame through a shaped flange. The design of the interface ring allows HFI integration inside the LFI as well as waveguide paths, while ensuring accurate alignment of the 4~K reference loads with the reference horns mounted on the FEMs.

%% file: 413_feed_horns.tex
The LFI feeds must have highly symmetric beams, low levels of side lobes ($-$35~dB), cross-polarisation ($-$30~dB) and return loss ($-$25~dB), as well as good control of the phase centre location \citep{2002ExA....14....1V}. Dual profiled conical corrugated horns have been designed to meet these  requirements, a solution that has the further advantage of high compactness and design flexibility. The profiles have a sine squared inner section, i.e., $R(z) \propto$sin$^2(z)$, and by an exponential outer section, $R(z) \propto $exp$(z)$.

\begin{figure}[tmb]
\begingroup
\newdimen\tblskip \tblskip=5pt
\centerline{{\bf Table~4.} Specifications of the feed horns.}
\nointerlineskip
\vskip -2mm
\footnotesize
\advance\baselineskip by 2pt 
\setbox\tablebox=\vbox{
   \newdimen\digitwidth 
   \setbox0=\hbox{\rm 0} 
   \digitwidth=\wd0 
   \catcode`*=\active 
   \def*{\kern\digitwidth}
   \newdimen\decimalwidth 
   \setbox0=\hbox{$.0$} 
   \decimalwidth=\wd0 
   \catcode`!=\active 
   \def!{\kern\decimalwidth}
\halign{\hbox to 1.7in{#\leaderfil}\tabskip=1em&
    \hfil#\hfil&
    \hfil#\hfil&
    \hfil#\hfil\tabskip=0pt\cr
\noalign{\doubleline}
\omit&30\,GHz&44\,GHz&70\,GHz\cr
\noalign{\vskip 3pt\hrule\vskip 5pt}
Band [GHz]&27--33&39.6--48.4&63--77\cr
Return Loss [dB]&$<$$-25$&$<$$-25$&$<$$-25$\cr
Insertion Loss @20\,K [dB]&$-$0.1&$-$0.1&$-$0.1\cr
Edge Taper @22\deg [dB]&30&30&25\cr
Sidelobes [dB]&$<$$-35$&$<$$-35$&$<$$-35$\cr
Cross polarization [dB]&$<$$-30$&$<$$-30$&$<$$-30$\cr
\noalign{\vskip 5pt\hrule\vskip 3pt}}}
\enndtable
\endgroup
\end{figure}

The detailed electromagnetic designs of the feeds were developed based on the entire optical configuration of the feed--telescope system. The control of the edge taper only required minimal changes on the feed aperture and overall feed sizes, so that an iterative design process could be carried out at system level. 

The evaluation of straylight effects in the optimisation process required extensive simulations (carried out with GRASP8 software) of the feed-telescope assembly for several different feed designs, edge tapers, and representative position in the focal plane \citep{2009_LFI_cal_M5}. A multi-GTD (geometrical theory of diffraction) approach was necessary since the effect of shields and multiple scatter needed to be included in the simulations. In Table~4 we report the main requirements and characteristics of the LFI feed horns. The details of the design, manufacturing, and testing of the LFI feed horns are discussed in \citet{2009_LFI_cal_O1}.

%% file: 414_omts.tex
The use of orthomode transducers (OMTs) allows the full power intercepted by the feed horns to be used by the LFI radiometers, and makes each receiver intrinsically sensitive to linear polarisation. The OMT splits the TE$_{11}$ propagation mode from the output circular waveguide of the feed horn into two orthogonal polarised components. Low insertion loss ($<0.15$dB) is needed to minimise impact on radiometer sensitivity. In addition, OMTs are critical components for achieving the ambitious wide bandwidth specification, especially when combined with the miniaturisation imposed by the focal plane arrangement.

\begin{figure}[tmb]
\begingroup
\newdimen\tblskip \tblskip=5pt
\centerline{{\bf Table~5.} Specifications of the OMTs.}
\nointerlineskip
\vskip -2mm
\footnotesize
\advance\baselineskip by 2pt 
\setbox\tablebox=\vbox{
   \newdimen\digitwidth 
   \setbox0=\hbox{\rm 0} 
   \digitwidth=\wd0 
   \catcode`*=\active 
   \def*{\kern\digitwidth}
   \newdimen\decimalwidth 
   \setbox0=\hbox{$.0$} 
   \decimalwidth=\wd0 
   \catcode`!=\active 
   \def!{\kern\decimalwidth}
\halign{\hbox to 1.7in{#\leaderfil}\tabskip=1em&
    \hfil#\hfil&
    \hfil#\hfil&
    \hfil#\hfil\tabskip=0pt\cr
\noalign{\doubleline}
\omit&30\,GHz&44\,GHz&70\,GHz\cr
\noalign{\vskip 3pt\hrule\vskip 5pt}
Band [GHz]&27--33&39.6--48.4&63--77\cr
Isolation [dB]&$<$$-40$&$<$$-40$&$<$$-40$\cr
Return Loss [dB]&$<$$-20$&$<$$-20$&$<$$-20$\cr
Insertion Loss @20\,K [dB]&$-$0.15&$-$0.15&$-$0.15\cr
Cross polarization [dB]&$<$$-25$&$<$$-20$&$<$$-20$\cr
\noalign{\vskip 5pt\hrule\vskip 3pt}}}
\enndtable
\endgroup
\end{figure}

These requirements made commercial OMTs inadequate for LFI, and a dedicated design development was carried out for these components \citep{2009_LFI_cal_O2}. An asymmetric design was selected, with a common polarisation section connected to the feed horn and to the main and side arms in which the two polarisations are separated. A modular design approach was developed, in which six different sections were identified, each corresponding to a specific electromagnetic function. While the basic configuration of the OMTs at different frequencies is scaled from a common design, some details are optimised depending on frequency, such as the location of the waveguide twist in the side arm or in the main arm. The main design specifications for the LFI orthomode transducers are given in Table~5. Detailed discussion of the design, manufacturing and unit-level testing of the LFI OMTs are reported in \citet{2009_LFI_cal_O2}.

%% file: 415_fems.tex
The performance of the LFI relies largely on its front-end modules (FEMs, see Table~6). Detailed descriptions of the 30 GHz and 44 GHz FEMs are given by \citet{2009_LFI_cal_R8}, and of the 70~GHz FEMs by \citet{2009_LFI_cal_R10}.

Each FEM accepts four input signals, two from the rectangular waveguide outputs of the OMT and two from the 4~K reference loads viewed by rectangular horns attached to the FEMs (Fig.~\ref{fig:rca_schematics_big}). 

\begin{figure}[h!]
    \begin{center}
        \includegraphics[width=7.5cm]{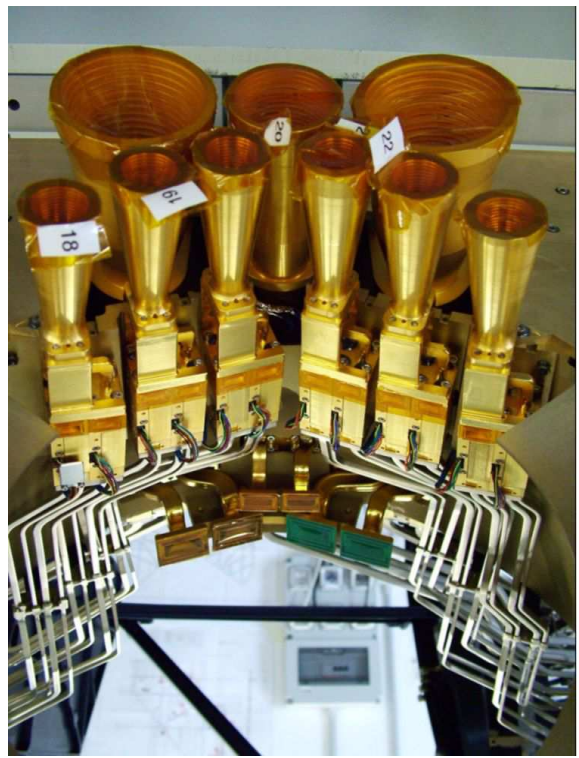}
    \end{center}
    \caption{Picture of nine of the eleven feed horns and associated OMTs, FEMs and waveguides before integration of the HFI front end. 
The six smaller feeds in the front supply the 70~GHz channels. In the back row, the two 30 GHz feeds flank one of the 44 GHz feeds. Visible are part of the twisted Cu waveguide sections. The reference horns are visible (protected by a kapton layer): for the 70~GHz FEMs they are embodied in the FEMs, while for the 30 and 44~GHz they are flared waveguide sections.}
    \label{fig:fems}
\end{figure}

In each half-FEM, the sky and reference input signals are connected to a hybrid coupler (or ``magic-T''). To minimise front-end losses, waveguide couplers were used, machined in the aluminium alloy FEM body and gold-plated. At 20~K the hybrid losses are estimated 0.1 to 0.2~dB. The first hybrid divides the signals between two waveguide outputs along which the low noise amplifiers and phase shifters are mounted. The internal waveguide design ensures that the phase is preserved at the input of the second hybrid. The signals are thus recombined at the FEM outputs as voltages proportional to the sky or reference load signal amplitude, depending on the state of the modulated phase switches.

At 70~GHz the large number of channels called for a highly modular FEM design, where each half-module could be easily dismounted and replaced. In addition, the elements hosting the low noise amplifiers (LNAs) and the phase shifters (the so called amplifier chain assembly, ACA, see Fig.~\ref{fig:rca_schematic_and_picture}) are built as separable units. This allowed flexibility during selection and testing, which proved extremely useful when replacement with a spare unit was required in an advanced stage of integration (see \citet{2009_LFI_cal_M3}). At 30 and 44~GHz, a multi-splitblock solution was devised to facilitate testing and integration, in which the four ACAs were mounted to end-plates and arranged in a mirror-image format.

\begin{figure*}[tmb]
\begin{center}
\begingroup
\newdimen\tblskip \tblskip=5pt
\centerline{{\bf Table~6.} Specifications of the front-end modules.}
\nointerlineskip
\vskip -3mm
\footnotesize
\advance\baselineskip by 2pt 
\setbox\tablebox=\vbox{
   \newdimen\digitwidth 
   \setbox0=\hbox{\rm 0} 
   \digitwidth=\wd0 
   \catcode`*=\active 
   \def*{\kern\digitwidth}
   \newdimen\signwidth 
   \setbox0=\hbox{+} 
   \signwidth=\wd0 
   \catcode`!=\active 
   \def!{\kern\signwidth}
\halign{\hbox to 3.8in{#\leaderfil}\tabskip=2em&
    \hfil#\hfil&
    \hfil#\hfil&
    \hfil#\hfil\tabskip=0pt\cr
\noalign{\doubleline}
\omit&30\,GHz&44\,GHz&70\,GHz\cr
\noalign{\vskip 3pt\hrule\vskip 5pt}
Band [GHz]&27--33&39.6--48.4&63-77\cr
Noise temperature over band [K]&8.6&14.1&25.7\cr
Gain (average over band) [dB]&30--33&30--33&30--33\cr
Gain variation with physical temperature [dB/K]&0.05&0.05&0.05\cr
Noise temperature variation with physical temperature [K/K]&0.8&0.8&0.8\cr
$1/f$ knee frequency [mHz]&$<$20&$<$20&$<$20\cr
Cross polarization [dB]&$<$$-35$&$<$$-35$&$<$$-35$\cr
Average power dissipation per FEM [mW]&31&31&24\cr
\noalign{\vskip 5pt\hrule\vskip 3pt}}}
\enndtable
\endgroup
\end{center}
\end{figure*}

\paragraph{LNAs.} In order to meet LFI requirements it was necessary to reach amplifier noise temperatures lower than previously achieved with multi-stage transistor amplifiers. We implemented state-of-the art cryogenic Indium Phosphide (InP) HEMT technology at all frequencies. Each front-end LNA must have a minimum of 30~dB of gain to reject back-end noise, which required 4 to 5 stage amplifiers. In addition to low noise, the InP technology enables very low power operation, which is essential to meet the requirements for heat load at 20~K. The amplifiers were selected and tuned for best operation at low drain voltages and for gain and phase match between paired radiometer legs, which is crucial for good balance. 

\begin{figure}[h!]
    \begin{center}
        \includegraphics[width=7.5cm]{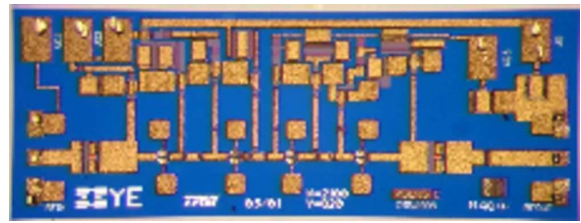}
    \end{center}
    \caption{A 4-stage Indium Phosphide HEMT MMIC low noise amplifier used in a flight model FEM at 70 GHz. The size of the MMIC is 2.1~mm $\times$ 0.8~mm.}
    \label{fig:mmic}
\end{figure}

The 70~GHz receivers are based on monolithic microwave integrated circuit (MMIC) semiconductors (Fig.~\ref{fig:mmic}) while at 30 and 44 GHz discrete HEMTs on a substrate (MIC technology) are used. To reach lowest possible noise temperatures, ultra-short gate devices are adopted. The final design used 0.1~$\mu$m gate length InP HEMTs manufactured by TRW (now Northrop Grumman) from the Cryogenic HEMT Optimisation Program (CHOP).

\begin{figure}[h!]
    \begin{center}
        \includegraphics[width=7.5cm]{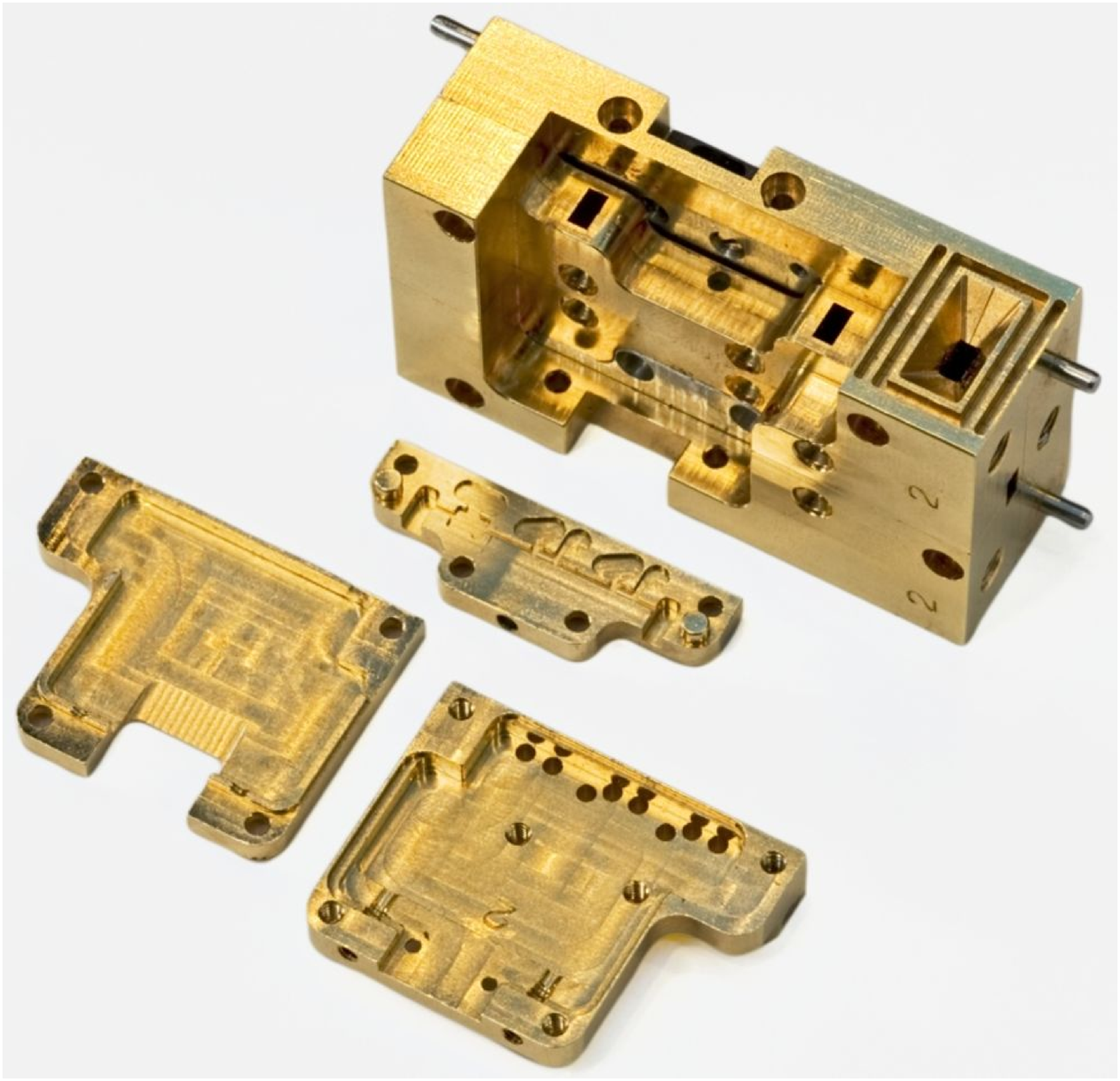}\\
        \includegraphics[width=7.5cm]{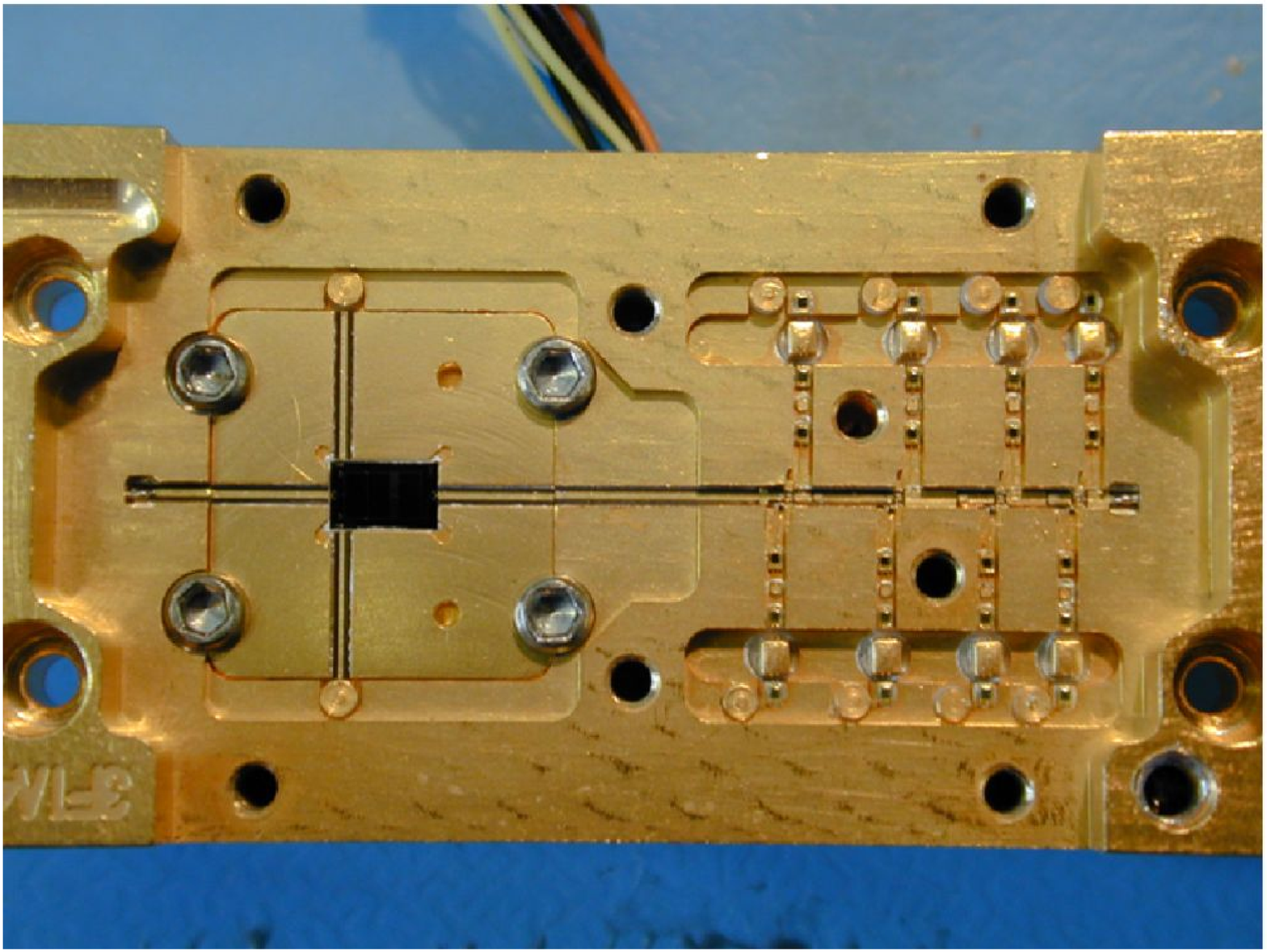}\\
    \end{center}
    \caption{Top: Structure of a 70~GHz half-FEM. On the right side of the module, one can see the small reference horn used to couple to the  4~K reference load surrounded by quarter-wave grooves. The parts supporting the amplifier chain assembly are dismounted and shown in the front. Bottom: picture of LNA and phase switch within a 30~GHz FEM. The RF channel incorporating the four transistors runs horizontally in this view. The phase switch is the black rectangular element on the left.}
    \label{fig:fem_lnas}
\end{figure}

\paragraph{Phase shifters.} After amplification the LNA output signals are applied to two identical phase shifters whose state is set by a digital control line modulated at 4~kHz. The signal is then conveyed via stripline to waveguide transitions to the second hybrid; the signal then 
passes to the interface with the interconnecting waveguide assembly to the BEMs. The phase switch design used at all frequencies is based on a double hybrid ring configuration \citep{2003_hoyland_espoo}. The switches, manufactured with InP PIN diodes, demonstrated excellent cryogenic performances for low 1/$f$ noise contribution and good 180$^\circ$ phase shift capability and amplitude balance.

\paragraph{Electrical connections.} Each of the 11 FEMs uses 16 to 20 low noise transistors and 8 phase switch diodes, all operated in cryo conditions. This sets demanding requirements in the design of the bias circuitry. The LNA biases are controlled by the data acquisition electronics (DAE) to obtain the required amplification and lowest noise operation. Different details in the design of the FEMs at each frequency minimise the number of supply lines. In the 30 and 44 GHz~FEMs potentiometers were used to common most of the control wires. The cryo-harness wiring that connects the FEMs to the power supply is $\sim$1.5~m in length (see Section \ref{sec:cryoharness}), which requires that each wire is terminated in the FEM with electrical protection devices to avoid sharp spikes.

\begin{figure}[h!]
    \begin{center}
        \includegraphics[width=7.5cm]{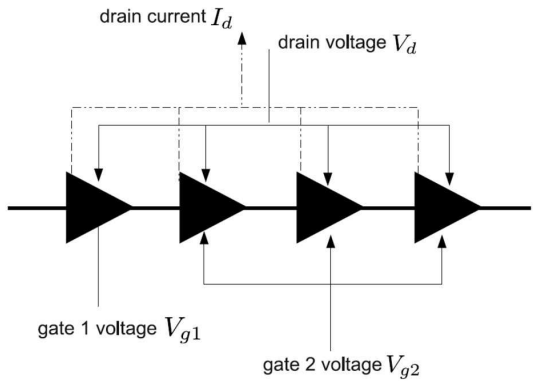}
    \end{center}
    \caption{DAE biasing to a front-end LNA. Each LNA is composed by four to five amplifier stages driven by a common drain voltage, a dedicated gate voltage to the first stage (most critical for noise performances) and a common gate voltage to the other stages.}
    \label{fig:bias_schematics}
\end{figure}

Proper tuning of the LNAs is critical for best performance. The front-end InP LNAs contain 4 stages of amplification at 30 and 70 GHz and 5 stages at 44 GHz. The LNAs are driven by three voltages: a common drain voltage ($V_d$); a gate voltage for the first stage ($V_{g_1}$); and a common gate voltage for the remaining stages ($V_{g_2}$) (Fig.~\ref{fig:bias_schematics}). The voltages $V_{g_1}$ and $V_{g_2}$ are programmable and are optimised in the tuning phase \citep{2009_LFI_cal_R7}. The total drain current, $I_d$ flowing in the ACA is measured and is available in the instrument housekeeping. 

%% file: 420_the_rl_system.tex
Blackbody loads provide stable internal signals for the pseudo-correlation receivers \citep{2009_LFI_cal_R1}. Cooling the loads as close as possible to the $\sim$3~K sky temperature minimises the radiometer knee frequency (Eq.~(\ref{eq:optimal_fk})) and reduces the susceptibility to thermal fluctuations and to other systematic effects. Requirements are thus derived for the loads absolute temperature, $T_\text{4K}<5$~K, as well as for the insertion loss of the reference horn, $L_\text{4K}<0.15$~dB.

Connecting the loads to the HFI 4~K stage imposes challenging thermo-mechanical requirements on the system. The loads must be thermally isolated from, but radiometrically matched to, the reference horns feeding the 20~K FEMs. Complete thermal decoupling is obtained by leaving a $\sim$1.5~mm gap between the loads and the reference horns. 

\begin{figure}[h!]
    \begin{center}
        \includegraphics[width=7.5cm]{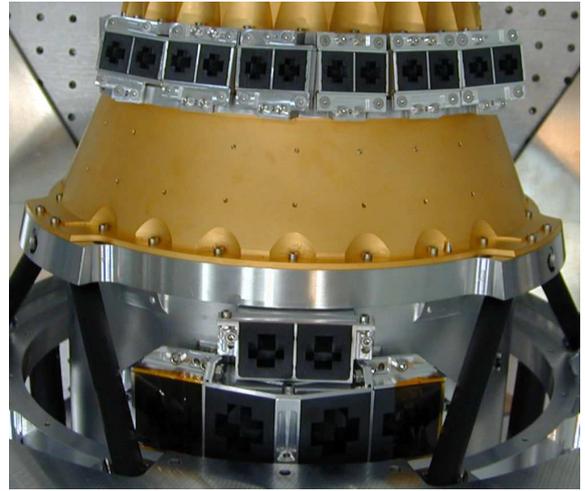}
    \end{center}
    \caption{Picture of the reference loads mounted on the HFI 4~K box. On the top is the series of loads serving the 70~GHz radiometers, while in the lower portion are the loads of the two 30 GHz FEMs and one of the 44~GHz FEMs. Note that each FEM is associated to two loads, each feeding a radiometer in the RCA.}
    \label{fig:reference_loads}
\end{figure}

Radiometric requirements for the loads were derived by analysis and tests \citep{2009_LFI_cal_R1}. The coupling between the reference horns and the loads is optimized to reduce reflectivity and to avoid straylight radiation leaking through the gap, with a required match at the horn aperture $>20$~dB. For the 70~GHz radiometers the reference horns are machined as part of the FEM body, while for the 30 and 44 GHz~they are fabricated as independent waveguide components and mounted on the FEM (Fig.~\ref{fig:fems}). This is because of the different paths required to reach the loads, whose location on the 4K box is constrained by the LFI-HFI interface.

The configuration of the loads is the result of a complex trade-off between RF performance and a number of constraints such as allowed mass, acceptable thermal load on the 4~K stage, and available volume. The latter is limited by the optical requirement of placing the sky horns close to the focal plane centre. In addition, the precise location and alignment of the loads on the HFI box need to follow the non-trivial orientation and arrangement of the FEMs, dictated by the optical requirements for angular resolution and polarisation. The final design comprises a front layer (made in ECR-110) shaped for optimal match with the reference horn radiation pattern, and a back layer (ECR-117) providing excellent absorption efficiency.

Requirements on thermal stability at the 4~K shield interface were set to $10\mu$K$\sqrt{s}$, i.e., at the same level as the HFI internal requirement. To maximise thermal stability, a PID (proportional, integral, and derivative) system was implemented on the HFI 4~K box \citep{2009_Lamarre_HFI}. The 70~GHz loads are located in the top of the HFI 4~K box, near the PID control system, while the 30 and 44~GHz loads are in the lower part (Fig.~\ref{fig:fems} and  Fig.~\ref{fig:reference_loads}). 
This resulted in a more stable signal for the LFI 70~GHz loads than for the 30 and 44~GHz ones. Simulations have shown that residual systematic effects on the maps at end-of-mission, after applying destriping algorithms \citep{Kei04}, are expected to be $\lsim 1.5 \mu$K at 30-44~GHz and $\lsim 0.2 \mu$K at 70~GHz.

%% file: 430_wgs.tex
A total of 44 waveguides connect the 20~K FEU and the 300~K BEU through a length of 1.5 to 1.9~m, depending on RCA. Conflicting constraints of thermal, electromagnetic and mechanical nature imposed challenging trade-offs in the design. The LFI waveguides must ensure good thermal isolation between the FEM and the BEM, while avoiding excessive attenuation of the signal. In addition, their mechanical structure must comply with the launch vibration loads. The asymmetric location of the FEMs in the focal plane, and the need to ensure integrability of the HFI in the LFI main frame as well as of the RAA on the spacecraft, impose complex routing with several twists and bends, which required a dedicated design for each individual waveguide.

\begin{figure}[h!]
    \begin{center}
        \includegraphics[width=7.5cm]{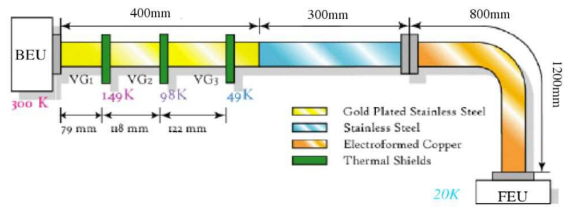}
    \end{center}
    \caption{Schematic of the LFI composite waveguide design, showing representative dimensions of the various sections as described in the text.}
    \label{fig:wg_design}
\end{figure}

\begin{figure*}[tmb]
\begin{center}
\begingroup
\newdimen\tblskip \tblskip=5pt
\centerline{{\bf Table~7.} RF requirements on the waveguides.}
\nointerlineskip
\vskip -3mm
\footnotesize
\advance\baselineskip by 2pt 
\setbox\tablebox=\vbox{
   \newdimen\digitwidth 
   \setbox0=\hbox{\rm 0} 
   \digitwidth=\wd0 
   \catcode`*=\active 
   \def*{\kern\digitwidth}
   \newdimen\signwidth 
   \setbox0=\hbox{+} 
   \signwidth=\wd0 
   \catcode`!=\active 
   \def!{\kern\signwidth}
\halign{\hbox to 2.2in{#\leaderfil}\tabskip=2em&
    \hfil#\hfil&
    \hfil#\hfil&
    \hfil#\hfil\tabskip=0pt\cr
\noalign{\doubleline}
\omit&30\,GHz&44\,GHz&70\,GHz\cr
\noalign{\vskip 3pt\hrule\vskip 5pt}
Insertion loss [dB]&$-$2.5&$-$3&$-$5\cr
Return loss [dB]&$<$25&$<$25&$<$25\cr
Isolation [dB]&$<$30&$<$30&$<$30\cr
Waveguide size&WR28&WR22&WR12\cr
Waveguide internal section [mm]&7.112$\times$3.556&5.690$\times$2.845&3.099$\times$1.549\cr
\noalign{\vskip 5pt\hrule\vskip 3pt}}}
\enndtable
\endgroup
\end{center}
\end{figure*}

A composite configuration was devised with two separated sections: a stainless steel (SS) straight section connecting to the BEMs, and a Copper (Cu) section incorporating all the twists and bends and connecting to the FEMs (Fig.~\ref{fig:wg_design}). The two sections are connected via custom-designed multiple flanges, each serving the four guides in each RCA. 

\begin{figure}[h!]
    \begin{center}
        \includegraphics[width=7.5cm]{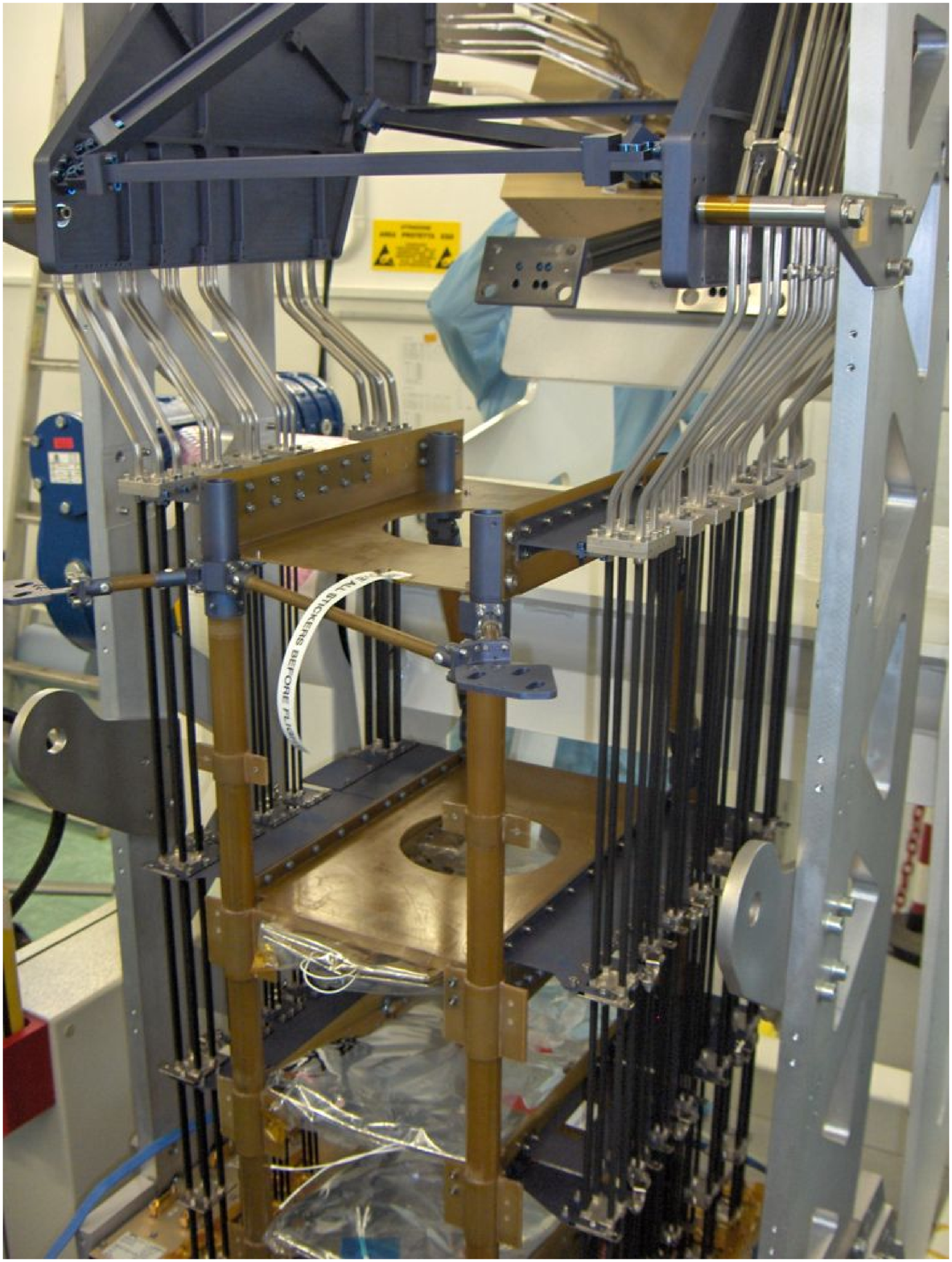}\\
\bigskip
        \includegraphics[width=7.5cm]{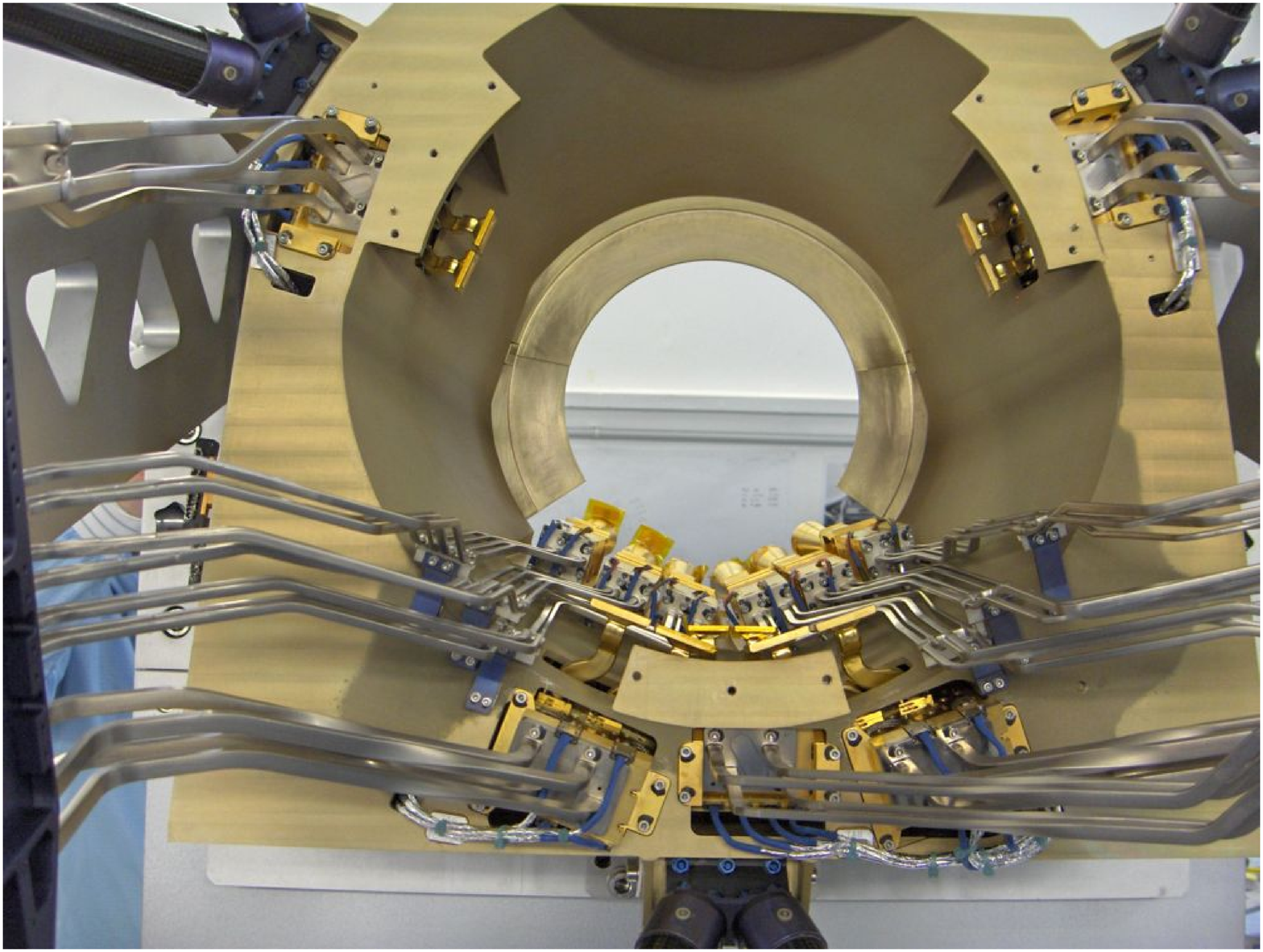}
    \end{center}
    \caption{Pictures of the LFI waveguides mounted on the RAA during the integration of the LFI flight model. Top: a view of the straight SS sections, black-painted on the outside and arranged in groups of four. On the upper part the Cu sections are connected through multiple flanges. Visible are also the upper and lower mechanical support structures. The three interface levels corresponding to the three V-grooves are also shown. Bottom: back view of the LFI front-end unit showing the twisted Cu sections connecting to the FEMs. The waveguide routing and central hole in the main frame are designed to interface with the HFI front-end 4~K box.}
    \label{fig:wg_picture}
\end{figure}

The SS sections support essentially the entire 20K--300K thermal gradient. All 44 SS waveguides have the same length (70 cm) and are gold-plated (2~$\mu$m thickness) in the first 40 cm near the BEM interface to minimise ohmic losses.  They are thermally sunk to the three V-grooves (Fig.~\ref{fig:wg_picture}) and their outer surfaces are black painted (Aeroglaze Z306) to optimise heat radiation. 

The Cu sections, of lengths 80 to 120~cm depending on RCA, operate at a nearly constant temperature of 20~K, and are individually designed based on optimization of return loss compatible with the required routing. Precise criteria for the curvature radii, twist length and mechanical tolerances were followed in the design \citep{2009_LFI_cal_O3}. Dynamical analysis showed the need for two dedicated mechanical support structures to ensure compliance with the vibration loads of the Ariane 5 launch (Fig.~\ref{fig:wg_picture}).

The selected design proved to meet simultaneously the heat load limits to the 20~K stage (250~mW for the bundle of 44 guides) and the insertion loss at a level of few dBs. A thermal model was developed to calculate the temperature profile along the waveguide and the final solution was found using an analytical model. 

\citet{2009_LFI_cal_O3} gives a full account of the manufacturing, qualification and challenging test plan of the LFI waveguides.

%% file: 441_bems.tex
The back-end modules (BEMs) are housed in the LFI back-end unit together with the DAE and are operated at room temperature $\sim$300~K. Each BEM has four branches, coupled in pairs (Fig.~\ref{fig:rca_schematics_big}). In each channel within the BEM the incoming signal is filtered by a band-pass filter, amplified by cascaded transistor amplifiers, detected by a detector diode, and DC-amplified. The BEM casing also incorporates bias and protection circuits and connectors. Room temperature noise figures $<$3~dB and an overall amplification of 20 to 25~dB, depending on RCA, are specified for the BEM channels. Amplifier and detector diode instabilities are efficiently removed by the 4~kHz phase switching, so that amplifier knee frequencies of order 100 Hz are acceptable.

\paragraph{Amplifiers.} The 30 and 44~GHz BEMs \citep{2009_LFI_cal_R9} use MMIC Gallium Arsenide (GaAs) amplifiers. Each LNA consists of two cascaded stages. The 30~GHz MMICs are commercial circuits using four stages of pseudomorphic HEMTs with an operating bandwidth from 24 to 36~GHz, 23~dB of gain and 3~dB noise figure. The 44~GHz MMICs were manufactured with a process employing a 0.2~$\mu$m gate length P-HEMT on GaAs. The 70~GHz BEMs \citep{2009_LFI_cal_R10} use the same type of Indium Phosphide MMIC amplifiers as in the FEMs. Although not required for performance or power dissipation constraints, this solution proved convenient in conjunction with the FEM development.

\paragraph{Band pass filters.} Band pass filters in the BEMs are used to define the bandwidth and to reject out-of-band parasitic signals. In the 30 and 44~GHz units the filter is based on a microstrip-coupled line structure which inherently provides bandpass characteristics. Waveguide filters are used at 70~GHz, where the wavelength-scale cavity has an acceptable size.

\paragraph{Detector diodes} The detector design at 30 and 44~GHz uses commercially available GaAs planar doped barrier Schottky diodes. The diodes are mounted with a coplanar-to-microstrip transition to facilitate on-wafer testing prior to integration in the BEM. As in the FEM design, at 70~GHz the filters and the amplifier-detector assemblies of paired channels were mounted on separable modules to offer greater flexibility in the testing and optimisation phases.

\paragraph{DC amplifiers} The detector diode is followed by a low noise DC-amp with a voltage gain adequate to the required analogue output voltage range for the DAE interface. In order to meet EMC requirements and grounding integrity the output voltages are provided as differential signals.

\begin{figure}[h!]
    \begin{center}
        \includegraphics[width=7.5cm]{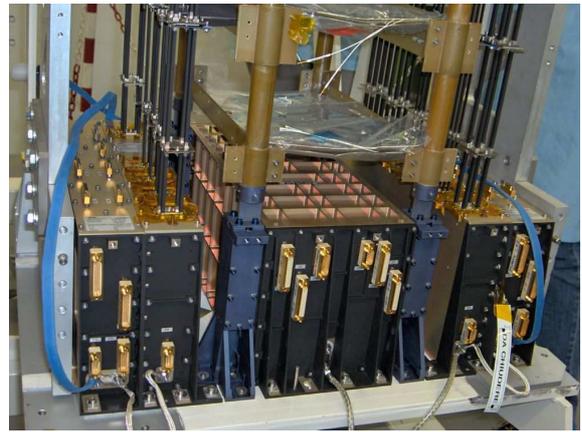}
    \end{center}
    \caption{Picture of the lower part of the LFI RAA during an advanced phase of the instrument flight model integration showing the LFI back-end unit. The two lateral trays hosting the radiometer BEMs are symmetrically disposed to the left and right sides of the DAE-BEU box. Visible are the lower part of the straight stainless steel waveguides.}
    \label{fig:beu}
\end{figure}

%% file: 442_dae.tex
The acquisition and conditioning of the science signals from the BEMs and of housekeeping data is performed by the data acquisition electronics. The DAE also provides power supply, conditioning and distribution to the RAA, and in particular DC biasing to the LNAs and phase switches in the FEMs, as well as to the BEM amplifiers. The DAE tags the acquired data using the information of its on-board time to ensure that correlation can be made on ground for proper pointing reconstruction. The raw data are then transmitted to the science processing unit for on-board processing.

As shown schematically in Fig.~\ref{fig:lfi_block_diagram}, the DAE functions are distributed into different sub-units. The ``DAE-BEU box'' and the ``lateral trays'' incorporate all the main functions, and are located in the back-end unit together with the radiometer BEMs (Fig.~\ref{fig:beu}). A separated ``DAE power box'' is interfaced with the spacecraft in order to receive the primary power supply and generate the secondary voltages needed to the DAE.

The DAE-BEU box is in charge of conditioning and acquiring the science data. The signals coming from the 44 detectors are integrated and held during the synchronous sampling and conversion. Science signals are digitised with 14-bit analogue-to-digital converters using a successive approximation conversion algorithm. There are 44 independent analogue acquisition chains, one for each detector arm. In order to optimise the analog signals from the BEMs to the ADC dynamic range, dedicated circuits remove an offset and then amplify the DC signal. Both offset and gain are programmable and are optimised as part of the instrument calibration process \citep{2009_LFI_cal_R7, 2009_LFI_cal_M3}. In the optimised configuration, the number of counts exercised by the radiometer noise varies from 10 to 450, depending on channel. Acquired data are converted into serial streams and automatically transferred to the signal processing unit in the REBA through synchronous serial links for processing and compression (Sect.~\ref{sec:reba}).

The DAE is also in charge of collecting and storing housekeeping data in a dedicated RAM. This information is retrieved by the REBA and organised in two dedicated packets with periods of 1 and 32 seconds respectively, depending on the needed monitoring frequency. Housekeeping parameters include current consumptions in the FEMs and temperature sensors, which are essential for trend analysis and systematic error tests.  These data are used extensively during the functionality checks of the radiometers.

The two DAE ``lateral trays'' contain the circuitry needed to provide the power supply to the RCAs, divided in four power groups (Fig.~\ref{fig:lfi_block_diagram}). These power supplies are independent from each other to minimise crosstalk and interference. The bias of each FEM and the controls for the phase switches are regulated to selectable voltage levels and filtered to achieve a minimum level of conducted noise. Particularly critical for the instrument performance is the optimal biasing of the FEM amplifiers. As described in Sect.~\ref{sec:fems}, the bias voltages of the first LNA stage and of the following stages are separately programmable.

%% file: 450_reba.tex
Downstream of the DAE, the LFI signals are digitally processed by the REBA (radiometer electronics box assembly), which also contains the power supply for LFI and the interface with the satellite SVM. The electronics hardware and on-board software are discussed by \citep{2009_LFI_cal_REBA}. The REBA is a fully redundant unit and it is internally separated in different sub-units as shown schematically in Fig.~\ref{fig:lfi_block_diagram}.

The signal processing unit (SPU) receives the raw digital science data from the DAE and performs on-board signal averaging, data compression and science telemetry packetisation. The need to reject 1/$f$ noise led to raw data sampling at 8192~Hz (122$\mu$s$/$sample), the LFI internal clock generator frequency. The clock drives synchronously the phase switches in the FEMs, the ADCs, and the on-board processor, which reconstructs the ordering of the acquired signals and synchronises it with the on--board time. Taking into account housekeeping and ancillary information, this corresponds to a data rate of $\sim$5.7 Mbps, or a factor of 100 higher than the allocated data rate for the instrument, 53.5~Kbps. Averaging the samples from sky and reference load signals to within the ''Nyquist" rate on the sky (3 bins per HPBW at each frequency) drastically reduces the data volume, leaving a compression requirement of a factor 2.4 (see Sect.~\ref{sec:electrical_interfaces}). The adopted algorithm implemented in the SPU relies on three-step processing of nearly loss--less compression which requires 5-parameter tuning to be optimised. The details of the LFI data compression strategy and end-to-end test results are discussed by \citet{2009_LFI_cal_D2}.

The main functions of the data processing unit (DPU) include monitoring and control of the RAA, instrument initialisation, error management, on-board time synchronisation, 
management of instrument operating modes, and control of the overall LFI data rate and data volume. Switching on and off the FEMs and BEMs as well as voltage adjustments are addressed by the DPU with a configuration that allows flexible setup commands. The DPU interface provides all commands for the DAE, while the SPU interface is in charge of retrieving
the fixed format raw data from the RCAs. Both the DPU and the SPU are based on a 18~MHz CPU. The link between the REBA and the DAE is implemented through IEEE 1355 interfaces and by means of data flag signals which ensure hardware and software synchronisation.

Finally, the data acquisition unit (DAU) is in charge of functions that are internal to the REBA and has no interfaces with the RAA. It converts the primary power received from the spacecraft to the secondary regulated voltages required by the REBA and it performs analog to digital conversion of REBA housekeeping data.

%% file: 510_20k_stage.tex
\begin{figure}[h!]
 \begin{center}
    \includegraphics[width=8.0cm]{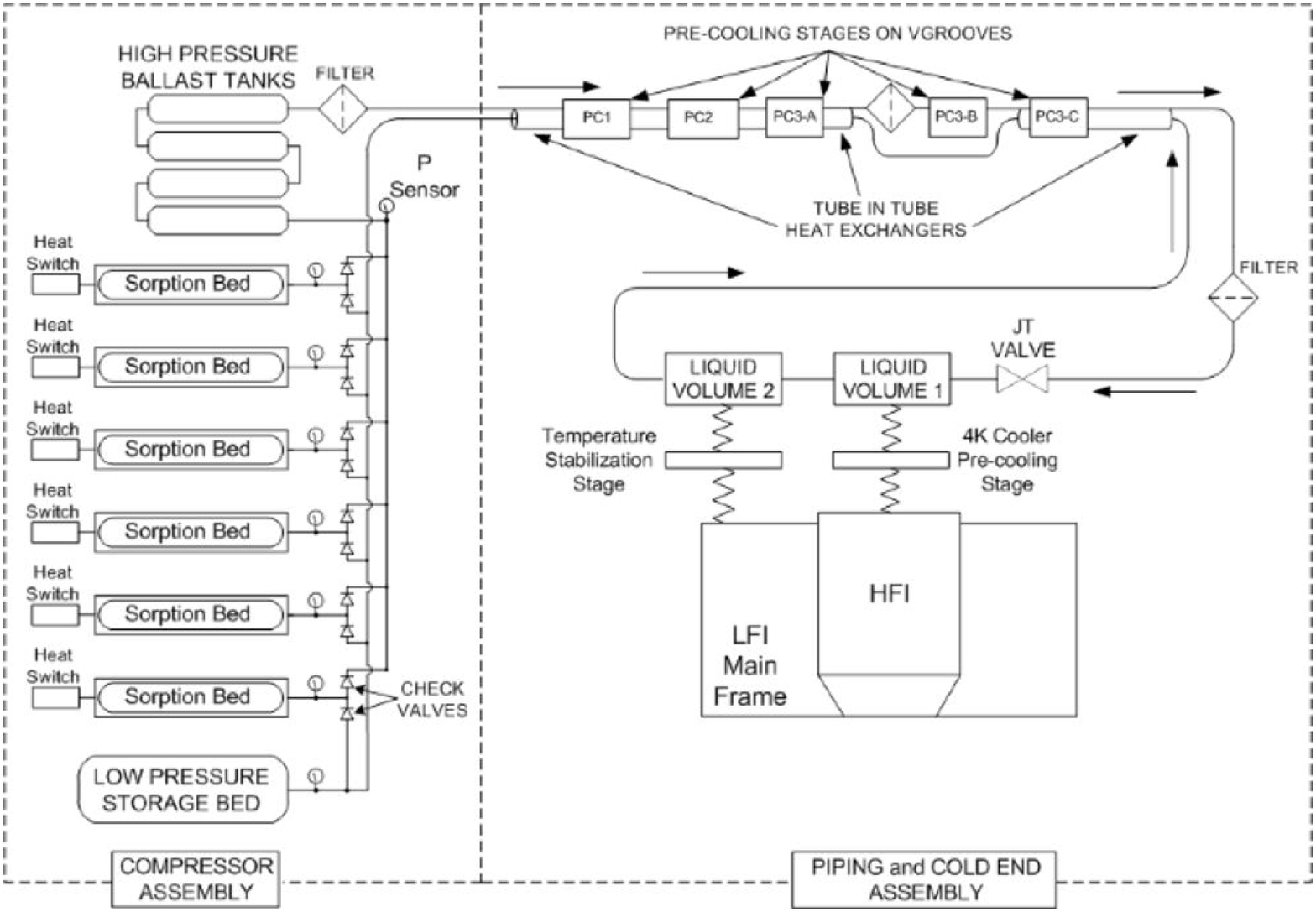}
 \end{center}
    \caption{Schematic of the 20K sorption cooler. The compressor assembly, located in the Planck service module, is shown in the left and comprises the six sorption beds, the low pressure storage bed, the high pressure tanks. The pre-cooling stages on the V-grooves, the J-T valve and the cold-end interfaces with the LFI and HFI instruments are shown on the right.}
    \label{fig:sorption_cooler}
\end{figure}

The LFI front-end is cooled to 20~K by a closed-cycle hydrogen sorption cryo-cooler \citep{Wade00, Bhandari04:planck_sorption_cooler,2009_LFI_cal_T6}, which also provides 18~K pre-cooling to the HFI (Fig.~\ref{fig:sorption_cooler}). The cooler provides $\sim$1~W of cooling power for the LFI FEU. The system operates by thermally cycling a set of compressors filled with La$_{1.0}$Ni$_{4.78}$Sn$_{0.22}$ powder alternately absorbing and desorbing H$_2$ gas as their temperature is cycled between $\sim$270~K and $\sim$450~K, thus providing the working fluid in a Joule-Thomson (JT) refrigerator. 

Heating of the sorbent beds is obtained by electrical resistance heaters, while cooling is achieved by thermally connecting the compressor element to a radiator at $\sim$270~K in the warm spacecraft. The hydrogen flow-lines are connected to the three V-groove radiators and passively pre-cooled to $<$50~K before reaching the 20~K JT expansion valve. 

In the complete system, six identical compressors are used, while a high-capacity storage sorbent bed is used as a gas reservoir in the low pressure line. At any time, one compressor is hot and desorbing to provide high pressure hydrogen gas in the range 30-50 atm, one compressor is heating up, one is cooling down, while the other three are cold and absorbing gas at $\sim$0.25 atm. This principle of operation ensures that no vibrations affect the detectors, a unique property of this kind of cooler which is very beneficial to Planck.

As a consequence of the cooler cycles the system exhibits two characteristic (controllable) time periods: $\tau_\text{bed}$, syncronous with each sorbent bed cycle (nominally $\tau_\text{bed} = 667$~s), and $\tau_\text{system}=6\times \tau_\text{bed}$, for the whole system cycle. 

Temperature fluctuations at the cold end are expected to modulate at these periods and may affect the LFI scientific performance both by direct coupling to the 20~K FPU and by fluctuations induced in the 4~K reference loads through the pre-cooling interface with HFI at 18~K. We analysed extensively this source of systematic effects in the design phase by propagating its effects to the map level \citep{mennella:sorption-cooler-requirements}. We derived stringent requirements ($\delta T<100$~mK peak-to-peak) on acceptable temperature fluctuations at the interfaces of the 20~K cooler with LFI (LVHX2). Testing at instrument level \citep{2009_LFI_cal_T3} and at system level have verified the design consistency.

%% file: 520_thermal_loads.tex
The limited cooling power of the sorption cooler imposes requirements on acceptable heat loads at 20~K. This includes power dissipated by the amplifiers and phase switches in the front-end, as well as parasitic loads from the waveguides, cryo-harness, and other passive elements. As discussed in Section \ref{sec:lfi_configuration_subsystems} these were strong drivers in the architecture of the radiometer chains and in the design of the waveguides and cryo-harness (Sections \ref{sec:wgs} and \ref{sec:cryoharness}). Table~8 summarises the heat load budget at 20~K for the LFI elements. The 299~mW allocated to the front-end modules has been split in an average dissipation of 31~mW per FEM at 30 and 44~GHz, and 24~mW per FEM at 70~GHz.


\begin{figure}[tmb]
\begingroup
\newdimen\tblskip \tblskip=5pt
\centerline{{\bf Table~8.} 20\,K heat load budget.}
\nointerlineskip
\footnotesize
\advance\baselineskip by 2pt 
\setbox\tablebox=\vbox{
   \newdimen\digitwidth 
   \setbox0=\hbox{\rm 0} 
   \digitwidth=\wd0 
   \catcode`*=\active 
   \def*{\kern\digitwidth}
   \newdimen\decimalwidth 
   \setbox0=\hbox{$.0$} 
   \decimalwidth=\wd0 
   \catcode`!=\active 
   \def!{\kern\decimalwidth}
\halign{\hbox to 1.7in{#\leaderfil}\tabskip=1em&
    \hfil#\hfil\tabskip=0pt\cr
\noalign{\doubleline}
\omit\hfil Item\hfil&P [mW]\cr
\noalign{\vskip 3pt\hrule\vskip 5pt}
FEMs&299\cr
Waveguides&249\cr
Cryoharness&**3\cr
Struts&*23\cr
Radiant coupling&**2\cr
\noalign{\vskip 5pt}
\bf TOTAL&\bf 576\cr
\noalign{\vskip 5pt\hrule\vskip 3pt}}}
\enndtable
\endgroup
\end{figure}

As part of the system thermal design, upper limits to heat loads on each of the three V-grooves were allocated to the LFI. Table~9 shows the estimated loads from the LFI compared to the budget allocations, showing that compliance has been achieved with ample margins.

\begin{figure}[tmb]
\begingroup
\newdimen\tblskip \tblskip=5pt
\centerline{{\bf Table~9.} LFI heat loads and allocations on V-grooves.}
\nointerlineskip
\footnotesize
\advance\baselineskip by 2pt 
\setbox\tablebox=\vbox{
   \newdimen\digitwidth 
   \setbox0=\hbox{\rm 0} 
   \digitwidth=\wd0 
   \catcode`*=\active 
   \def*{\kern\digitwidth}
   \newdimen\decimalwidth 
   \setbox0=\hbox{$.0$} 
   \decimalwidth=\wd0 
   \catcode`!=\active 
   \def!{\kern\decimalwidth}
\halign{\hbox to 1.1in{#\leaderfil}\tabskip=1em&
    \hfil#\hfil&
    \hfil#\hfil&
    \hfil#\hfil\tabskip=0pt\cr
\noalign{\doubleline}
\omit&$T$&$P_{\rm estimated}$&$P_{\rm allocated}$\cr
\omit&[K]&[mW]&[mW]\cr
\noalign{\vskip 3pt\hrule\vskip 5pt}
VG3&*51.4&*463&*710\cr
VG2&106.2&*337&*560\cr
VG1&166.2&2939&5370\cr
\noalign{\vskip 5pt\hrule\vskip 3pt}}}
\enndtable
\endgroup
\end{figure}

%% file: 530_temperature_sensors.tex
Temperature sensors are placed in strategic locations of the instrument flight model in order to monitor temperature values and fluctuations during both ground calibration \citep{2009_LFI_cal_T2, 2009_LFI_cal_T3} and in-flight operation. 

\begin{figure}[h!]
    \begin{center}
        \includegraphics[width=8.5cm]{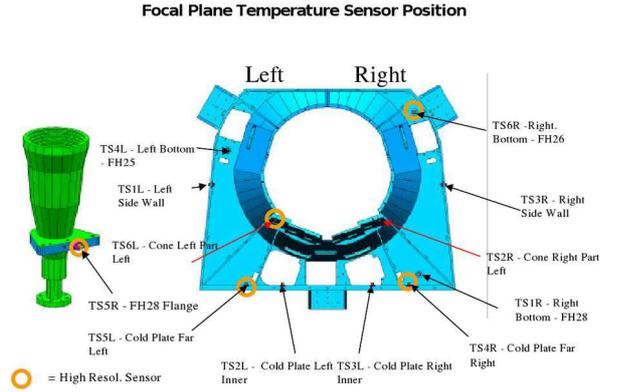}
    \end{center}
    \caption{Temperature sensors in the LFI front end unit. Circles indicate high sensitivity sensors.}
    \label{fig:feu_temperature_sensors}
\end{figure}

Fig.~\ref{fig:feu_temperature_sensors} shows the 12 sensors in the focal plane unit, five of which have higher sensitivity and lower dynamic range (14 to 26.5~K) to adequately trace temperature fluctuations. They are Lakeshore Silicon diodes DT670 and the associated readout electronics leads to a typical sensitivity of 0.9 mK at 20~K. Averaging of multiple readings allows to increase the resolution below the quantization limit at the relevant fluctuations timescale. One of the sensors (TS5R) has been placed on the flange of a 30~GHz feedhorn (RCA28) to directly monitor front-end stability. The interface to the sorption cooler cold end (LVHX2) is also monitored with sensors both on the LFI side and on the sorption cooler side. For analysis of in-flight data, additional temperature information will be used from sensors belonging to the HFI, the telescope and the spacecraft.

%% file: 610_cryo_harness.tex
\begin{figure*}
    \begin{center}
        \includegraphics[width=16cm]{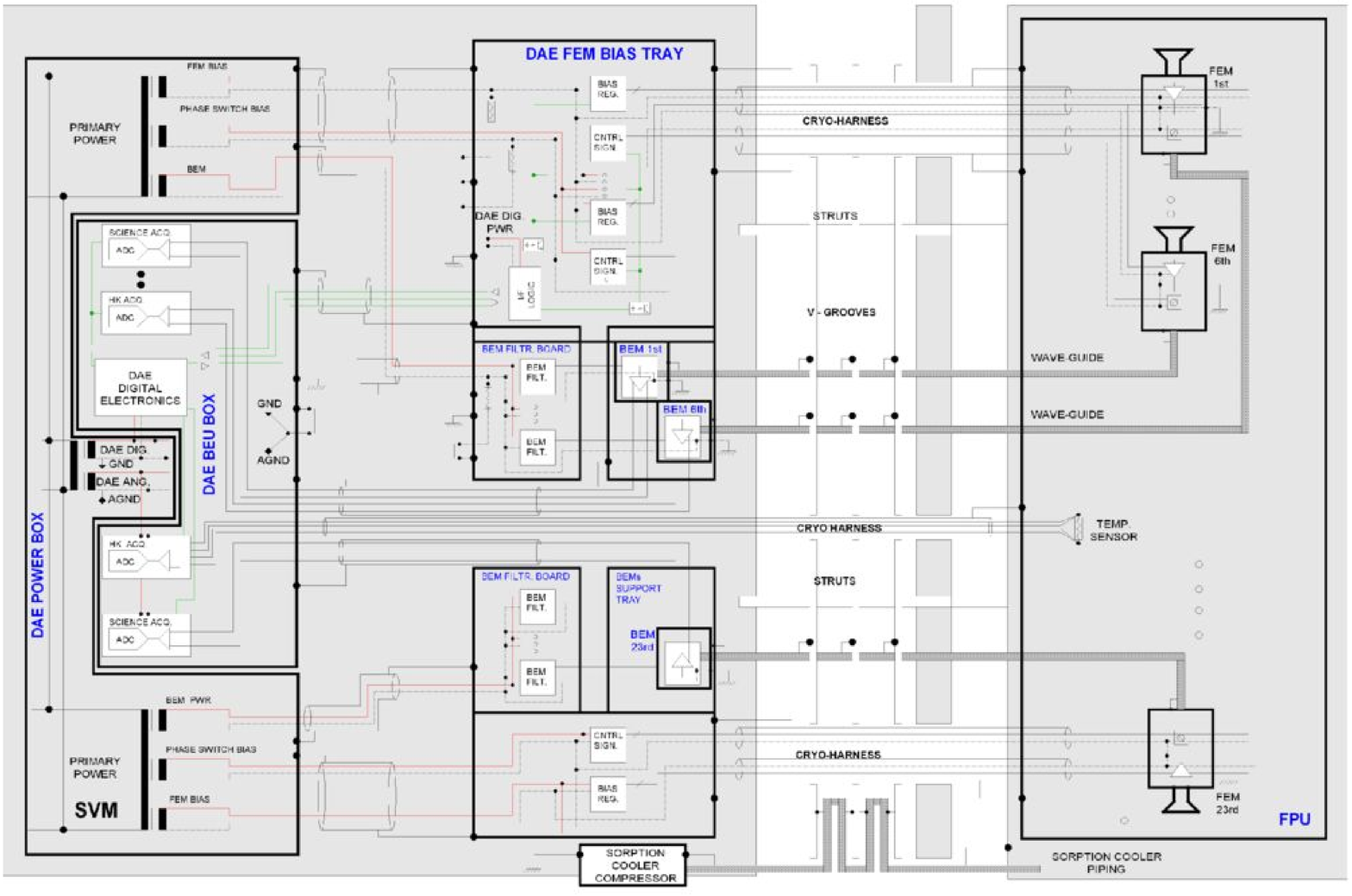}
    \end{center}
    \caption{Schematic of the grounding scheme of LFI}
    \label{fig:grounding_scheme}
\end{figure*}

\begin{figure*}[tmb]
\begin{center}
\begingroup
\newdimen\tblskip \tblskip=5pt
\centerline{{\bf Table~10.} Main characteristics and specifications of the LFI cryoharness.}
\nointerlineskip
\vskip -3mm
\footnotesize
\advance\baselineskip by 2pt 
\setbox\tablebox=\vbox{
   \newdimen\digitwidth
   \setbox0=\hbox{\rm 0}
   \digitwidth=\wd0
   \catcode`*=\active
   \def*{\kern\digitwidth}
   \newdimen\signwidth
   \setbox0=\hbox{+}
   \signwidth=\wd0
   \catcode`!=\active
   \def!{\kern\signwidth}
\halign{\hbox to 1.5in{#\leaderfil}\tabskip=2em&
    \hfil#\hfil\tabskip=1em&
    \hfil#\hfil&
    \hfil#\hfil&
    \hfil#\hfil\tabskip=2em&
    \hfil#\hfil\tabskip=1em&
    \hfil#\hfil&
    \hfil#\hfil&
    \hfil#\hfil\tabskip=0pt\cr
\noalign{\doubleline}
\omit&\multispan4\hfil R{\tfs EQUIREMENT}\hfil&\multispan4\hfil D{\tfs ESIGN} S{\tfs OLUTION}\hfil\cr
\noalign{\vskip -3pt}
\omit&\multispan4\hrulefill&\multispan4\hrulefill\cr
\omit&&$I_{\rm max}$ 20\,K&$I_{\rm max}$ 300\,K&$R$&&Diameter&$R$&$P$ at FPU\cr
\omit\hfil I{\tfs TEM}\hfil&$N$&[mA]&[mA]&$\Omega$&Material&[AWG, mm]&$\Omega$&[mW]\cr
\noalign{\vskip 3pt\hrule\vskip 5pt}
HEMT GND           &*11&*40&200&**$<$1&Copper  &38 (0.1)*&**0.6&*6.32\cr
HEMT Drain         &*44&*10&*50&**$<$5&Nickel  &38 (0.1)*&**2.2&11.37\cr
HEMT Gate          &*88&100&100&$<$200&Manganin&40 (0.08)&193.4&*1.86\cr
Phase Switch GND   &*11&**4&**4&*$<$10&Titanium&32 (0.2)*&**5.2&*1.55\cr
Phase Switch       &*88&**1&**1&*$<$50&Titanium&38 (0.1)*&*21.0&*5.88\cr
Shield             &$\ldots$&$\ldots$&$\ldots$&$\ldots$&Al + Kapton&100\,nm&$\ldots$&*2.28\cr
Temperature Sensors&*48&*40&*40&$<$200&Manganin&40 (0.08)&139.3&*0.93\cr
\noalign{\vskip 5pt}
\bf TOTAL       &\bf 290&&&&&&&\bf30.20\cr
\noalign{\vskip 5pt\hrule\vskip 3pt}}}
\enndtable
\endgroup
\end{center}
\end{figure*}

A further challenging element in the LFI design is the electrical connection from the DAE power supply to the cryogenic front-end (cryoharness). Each FEM needs 22 bias lines for biasing the LNAs and phase switches. When including temperature sensor wires, a total of 290 lines have to be routed from the 300K electronics to the 20~K FPU along a path of $\sim$2.2~m. The lines need to transport currents ranging from a few $\mu$A (for temperature sensors) up to 200~mA. The stability needed in the bias of the cryogenic LNAs calls for high immunity to external noise and disturbances, i.e., efficient electrical shielding. On the other hand, heat transport to 20~K needs to be kept at the few mW level. Furthermore, to ensure operability of LFI at room temperature (a tremendous advantage in the integration and test process), the harness was required to be compatible with operation at 300~K.

\begin{figure}[h!]
    \begin{center}
        \includegraphics[width=8.5cm]{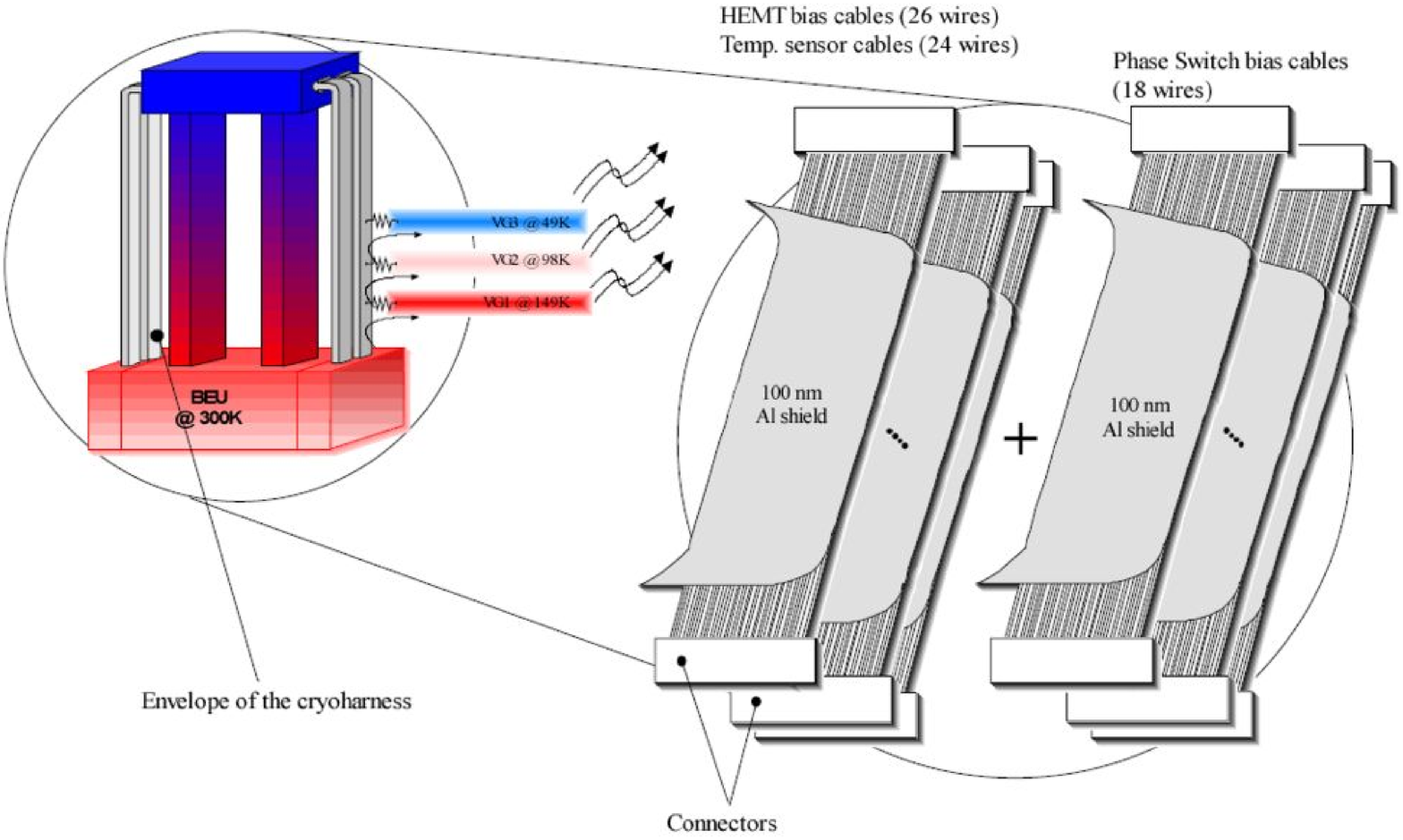}
    \end{center}
    \caption{Schematic of the cryoharness serving HEMT biasing, phase switch biasing and temperature sensors. Heat loads on the 20~K stage are minimised by intercepting heat with the V-grooves.}
    \label{fig:cryo_harness}
\end{figure}

In Table~10 we show the main characteristics of the implemented cable design  \citep{leute2003}, while Fig.~\ref{fig:cryo_harness} is a schematic of the cryoharness configuration and routing. The cryo-harness was mounted as part of the RAA and integrated before delivery, so that it could be kept integrated.

%% file: 620_emc.tex
Great effort has been made to ensure a highly stable electrical environment. The grounding scheme (Fig.~\ref{fig:grounding_scheme}) has been optimised to ensure maximum protection of the bias lines to the front-end cryogenic LNAs and phase switches. Unwanted fast voltage transient across the input biases can damage the InP--based HEMT junctions; furthermore, the front-end module performance requires very low noise in the received bias. 

The 20K LFI focal plane is about 2m away from the warm back-end unit, where the LNA biases are generated. Because InP technology requires the LNA substrates to be connected to the ground, in order to avoid ground loops, the only grounding reference for the instrument is on the focal-plane and the whole back-end electronics is referred to chassis at the radiometer's front-end.

The EMC design and verification approach has been performed incrementally and based on analysis and testing at component or sub-assembly level. Table~11 lists all the internal frequencies of the LFI instrument and sorption cooler. Whenever relevant, these have been monitored in the test campaign as potential sources of RF disturbances, both within LFI and towards the HFI detectors. In fact, a critical aspect in the design was to ensure mutual compatibility between the two instruments, which are in mechanical contact at the FPU interface. This issue was analysed in detail by both instrument teams, however, a hardware verification was possible only at FM system level during the cryogenic performance test campaign performed at CSL in summer 2008. The results confirmed excellent compatibility between the two instruments.

\begin{figure}[tmb]
\begingroup
\newdimen\tblskip \tblskip=5pt
\centerline{{\bf Table~11.} LFI Characteristic Internal Frequencies}
\nointerlineskip
\vskip -3mm
\footnotesize
\advance\baselineskip by 2pt 
\setbox\tablebox=\vbox{
   \newdimen\digitwidth 
   \setbox0=\hbox{\rm 0} 
   \digitwidth=\wd0 
   \catcode`*=\active 
   \def*{\kern\digitwidth}
   \newdimen\signwidth 
   \setbox0=\hbox{+} 
   \signwidth=\wd0 
   \catcode`!=\active 
   \def!{\kern\signwidth}
\halign{\hbox to 0.75in{#\leaderfil}\tabskip=1em&
    \vtop{\hsize=1.3in\hangindent=1em\hangafter=1\noindent\strut#\strut\par}&
    #\hfil\tabskip=0pt\cr
\noalign{\doubleline}
\omit\hfil$\nu$\hfil&\omit\hfil Origin\hfil&\omit\hfil Unit\hfil\cr
\noalign{\vskip 3pt\hrule\vskip 5pt}
1\,Hz     &Housekeeping acquisition frequency&DAE BEU\cr
1\,Hz     &Synchronisation signal&DAE BEU, REBA\cr
10\,Hz    &Internal timer&SCS\cr
1\,kHz    &Locking clocks&SCS\cr
4096\,Hz  &Phase Switch&FEM, DAE BEU\cr
100\,kHz  &5\,V \& 12\,V DC/DC&SCS\cr
131072\,Hz&DC/DC converters&DAE Power box\cr
131072\,Hz&On-board clock signal&DAE BEU, REBA\cr
131072\,Hz&LOBT clock&SCS\cr
200\,kHz  &12\,V DC/DC&SCS\cr
1\,MHz    &Command link from the BEU box&DAE power box\cr
1\,MHz    &Internal transfer of digital data&DAE BEU, REBA\cr
8\,MHz    &ADC clock&SCS\cr
10/80\,MHz&1355 serial data digital interface&DAE BEU, REBA\cr
16\,MHz   &DSP processor clock&SCS\cr
17.46\,MHz&Clock frequency of the DSP&REBA\cr
20\,MHz   &Sequencer internal clock&DAE BEU\cr
\noalign{\vskip 5pt\hrule\vskip 3pt}}}
\enndtable
\endgroup
\end{figure}

%% file: 630_data_rate.tex
The choice of an L$_2$ orbit for Planck induces stringent requirements on the rate of data transmission to the ground. Both the sky and reference load samples will be transmitted to the ground, so that full data reduction can be performed at the LFI Data Processing Centre (DPC). After sample-averaging and data compression (by a factor of 2.4 for at least the 95\% of the packets) performed in the REBA SPU (Sect.~\ref{sec:reba}), the science data volume is 36.12~Kbps, increased to 37.88~Kbps by packeting overheads. An additional contribution, up to 5.06~Kbps, comes from the so-called ``calibration channel'': for diagnostic purposes, one LFI channel at a time will be transmitted to the ground without compression. Adding 2.57~Kbps of housekeeping leads to a total budget of 45.41~Kbps for LFI, well within the allocated 53.5~Kbps (see Table~12). It is critical that the (average) 2.4 compression factor be achieved with an essentially lossless process, which requires careful optimisation of the parameters that control the on-board compression algorithm in the SPU \citep{2009_LFI_cal_D2}. After telemetry transmission the data will be treated through LFI DPC ``Level 1'' \citep{2009_LFI_cal_D3} for real-time assessment, housekeeping monitoring, data de-compression. Then the time-order information (TOI) will be generated and processed by the successive analysis steps in the DPC pipeline.


\begin{figure}[tmb]
\begingroup
\newdimen\tblskip \tblskip=5pt
\centerline{{\bf Table~12.} LFI data rate summary.}
\nointerlineskip
\vskip -2mm
\footnotesize
\advance\baselineskip by 2pt 
\setbox\tablebox=\vbox{
   \newdimen\digitwidth 
   \setbox0=\hbox{\rm 0} 
   \digitwidth=\wd0 
   \catcode`*=\active 
   \def*{\kern\digitwidth}
   \newdimen\decimalwidth 
   \setbox0=\hbox{$.0$} 
   \decimalwidth=\wd0 
   \catcode`!=\active 
   \def!{\kern\decimalwidth}
\halign{\hbox to 1.7in{#\leaderfil}\tabskip=1em&
    \hfil#\hfil&
    \hfil#\hfil&
    \hfil#\hfil\tabskip=0pt\cr
\noalign{\doubleline}
\omit&30\,GHz&44\,GHz&70\,GHz\cr
\noalign{\vskip 3pt\hrule\vskip 5pt}
Number of detectors           &8&12&24\cr
Angular resolution (nominal)            &33\rlap{\arcm}&24\rlap{\arcm}&14\rlap{\arcm}\cr
Beam crossing time [ms]       &92&64&39\cr
Sampling rate [Hz]            &32.51&46.55&78.77\cr
Science data rate [Kbps]      &8.32&17.87&60.49\cr
\noalign{\vskip 3pt\hrule\vskip 3pt}
Total science data rate       &&86.69\,Kbps\cr
\hglue 2.6em after compression &&36.12\,Kbps\cr
Total LFI data rate           &&45.41\,Kbps\cr
\noalign{\vskip 5pt\hrule\vskip 3pt}}}
\enndtable
\endgroup
\end{figure}

%% file: 700_optical_interfaces.tex
The optimisation of the optical interface between the combined LFI-HFI focal plane and the Planck telescope was coordinated
throughout the various development phases of the project. Rejection of systematic effects arising from non-ideal optical coupling has been a major design driver for LFI \citep{2000ApL&C..37..151M, 2009_LFI_cal_O1}. Minimisation of main beam ellipticity and distortion, particularly relevant for the off-axis LFI feeds, has been a key element in the optical design \citep{burigana98, 2009_LFI_cal_M5}. An upper limit of $<1$~$\mu$K (rms) to straylight contamination from various sources was set as a design criterion. 
Far sidelobe effects were simulated for the Galactic foregrounds (diffuse dust, free-free, and synchrotron emission, and HII regions) \citep{burigana01}, as well as for solar system sources. 
The full beam pattern was calculated using a combination of Physical Optics (PO), Physical Theory of Diffraction (PTD) and Multi-recflector Geometrical Theory of Diffraction (MrGTD) by considering radiation scattered by both reflectors, as well as reflection and diffraction effects on the baffle. 
The final LFI optical design is discussed by \citet{2009_LFI_cal_M5} (see \citet{2009_Maffei_HFI} for the analogous process for HFI). 
In particular, the LFI optical optimisation allowed to design the 70 GHz feeds as to meet straylight requirements and reach an angular resolution $\sim$13$'$ for most 70 GHz channels, thus improving over the requirements value (14$'$, Table~1).

Emission originating within the Planck spacecraft and coupling directly into the LFI beams (``internal straylight'') was also considered in the optical design as a potential source of systematic effect. An overall upper limit of 1${\mu}$K was set for internal straylight and a breakdown of contributions from various optical elements of the payload module (baffle, reflectors structures, third V-groove) was carried out. Simulations have shown compliance with the allocated budget.

Knowledge of the microwave transmission of the LFI channels is an essential element for the extraction of polarisation information 
\citep{2009_LFI_polarisation_M6} and for component separation. The band shapes of each RCA channel have been evaluated from measurements at single unit level, then combined with a dedicated software model, and finally verified with end-to-end testing as part of the RCA cryogenic test campaign \citep{2009_LFI_cal_R3}.

Alignment requirements on the FPU relative to the telescope were developed taking into account the thermo-elastic effects of the
cooldown to 50~K of the LFI struts. The driving requirements were set by the HFI optical alignment which were more stringent due to the shorter wavelengths. For the LFI, the internal alignment requirements between FEU and BEU of $\pm$2~mm required careful design of the waveguide support structures.

%% file: 900_conclusions.tex
The Planck scientific objectives call for full-sky maps with sensitivity $\Delta T/T\simeq 2\times 10^{-6}$ per $\Delta\theta\simeq 10'$ pixel. The combination of LFI and HFI covers the spectral range 30 to 850~GHz, to allow precise removal of non-cosmological emissions. The two instruments use widely different technologies and will be affected differently by different sources of systematic effects. This unique feature of Planck provides a powerful tool to identify and remove systematic effects. The Planck-LFI covers three frequency bands centred at 30, 44 and 70~GHz. The LFI is sensitive to polarisation in all channels, a characteristic of coherent detectors which does not call for any additional component or system compromise. The 70~GHz channel is near the minimum of the foreground emission, thus probing the cleanest cosmological window with an angular resolution of 13$'$. The 30 and 44~GHz channels are sensitive to the cosmological signal but also are sensitive to synchrotron, free-free and anomalous dust diffuse radiation from the Galaxy. Thus they will serve as cosmological and foreground monitors in the Planck observations. 


\begin{figure}[tmb]
\begingroup
\newdimen\tblskip \tblskip=5pt
\vbox{\hsize=88mm{\bf Table~13.} Principal requirements and design solutions in LFI.}
\nointerlineskip
\vskip -2mm
\footnotesize
\advance\baselineskip by 2pt 
\setbox\tablebox=\vbox{
   \newdimen\digitwidth 
   \setbox0=\hbox{\rm 0} 
   \digitwidth=\wd0 
   \catcode`*=\active 
   \def*{\kern\digitwidth}
   \newdimen\decimalwidth 
   \setbox0=\hbox{$.0$} 
   \decimalwidth=\wd0 
   \catcode`!=\active 
   \def!{\kern\decimalwidth}
\halign{\vtop{\hsize=30mm\hangafter=1\hangindent=1em\noindent\strut#\strut\par}
        \tabskip=1.4em&
   \vtop{\hsize=50mm\hangafter=1\hangindent=1em\noindent\strut#\strut\par}\tabskip=0pt\cr
\noalign{\doubleline}
\omit\hfil Requirement/\hfil\cr
\omit\hfil Constraint\hfil&\omit\hfil Design solution\hfil\cr
\noalign{\vskip 3pt\hrule\vskip 5pt}
High sensitivity&Cryogenically cooled ($\sim$20\,K) HEMT amplifiers.\cr
Low residual $1/f$, immunity from receiver systematics&Pseudo-correlation differential design.  Cryogenic reference load ($\sim$4\,K).  Offset removal by gain modulation factor in post-processing.  Fast switching (4\,KHz) of sky and reference signal to suppress backend $1/f$ noise.\cr
Single telescope&``Internal'' reference load.\cr
Modularity, cryo testing.&Phase switch in frontend modules.\cr
Low power dissipation on 20\,K stage.&Two amplification stages (cold frontend, warm backend).  Low loss and thermal conductivity interconnecting waveguides.\cr
Waveguide mechanical routing.&Phase switch and second hybrid in the frontend (avoids need of phase-matched waveguides.)\cr
\noalign{\vskip 5pt\hrule\vskip 3pt}}}
\enndtable
\endgroup
\end{figure}

The LFI design required several challenging trade offs involving thermal, mechanical, electrical and optical aspects (Table~13). The cryogenic front-end receivers, required for high sensitivity, dominate the instrument performance and their interface with HFI is a major driver of the instrument configuration. The combination of the pseudo-correlation scheme and of the 4~KHz switching of the phase shifters in the FEM allow us to obtain excellent stability while maintaining a highly modular design. Stringent requirements on noise temperature, 1/$f$ noise, thermal and electrical stability, bandwidth, polarisation isolation, and parasitic heat loads were key elements in the design. A further key driver in the LFI design has been the control of systematic effects, which has also been a central part of the LFI calibration plan and test campaigns, both on-ground and in-flight. 
The functionality and performance of LFI was tested at various stages of development and integration (component level, unit level, RCA, instrument and satellite level) and will be measured again in flight. The achieved performances based on ground testing are described by \citet{2009_LFI_cal_M3} and \citet{2009_LFI_cal_M4} and are generally in line with the design expectations.